\pdfoutput=1

\documentclass[12pt,nohyper,twosided]{JHEP3}
\usepackage{epsfig}
\usepackage{amssymb}
\usepackage{amsmath}
\usepackage{mathrsfs}

\newcommand{\crowS}[7]{$#1$&$\mathbf{#2}$&\ifnum#3<0$\!\!$#3\else#3\fi&\ifnum#4<0$\!\!$#4\else#4\fi&\ifnum#5<0$\!\!$#5\else#5\fi&\ifnum#6<0$\!\!$#6\else#6\fi&$#7=0$\\}
\newcommand{\crowF}[6]{$#1$&$\mathbf{#2}$&\ifnum#3<0$\!\!$#3\else#3\fi&\ifnum#4<0$\!\!$#4\else#4\fi&\ifnum#5<0$\!\!$#5\else#5\fi&$#6$\\}

\renewcommand{\bar}[1]{\overline{#1}}

\title{{\LARGE  Local Models in F-Theory and M-Theory with Three Generations}}
\author{Jacob L. Bourjaily\\Department of Physics, Princeton University, Princeton, NJ 08544, and\\School of Natural Sciences, Institute for Advanced Study, Princeton, NJ 08540
\\\email{jbourj@sns.ias.edu}}
\preprint{January 2009}

\abstract{We describe a general framework that can be used to geometrically engineer local, phenomenological models in F-theory and M-theory based on ALE-fibrations, and we present several concrete examples of such models that feature three generations of matter with semi-realistic phenomenology. We show that the geometric structures required for generating interactions---triple-intersections of matter-curves in F-theory and supersymmetric three-cycles supporting multiple conical singularities in M-theory---are generic in such ALE-fibred manifolds, and that they can be understood in correspondence with one another. The models we can construct in this way are strictly limited in complexity by the maximality of the $\widehat{E_8}$-ALE space, but turn out to be just complex enough to accommodate some of the most realistic string models to date.}
\begin{document}

\tableofcontents

\newpage
%
%
%
%
%
%
%
%
%
%
%
%
\section{Spiritus Movens}\vspace{-0.4cm}
There has been a great deal of recent interest in the possibility of geometrically engineering local, phenomenological models in F-theory and M-theory \cite{Beasley:2008dc,Beasley:2008kw,Heckman:2008ads,Heckman:2008rb,Heckman:2008qa,Bourjaily:2007vw,Bourjaily:2007vx,Bourjaily:2007kv,Bourjaily:2008ji,Donagi:2008ca,Marsano:2008jq,Marsano:2008py,Jiang:2008yf,Heckman:2008es}. In the framework of geometrical engineering, phenomenological data such as gauged and discrete symmetries, the spectrum of massless charged matter, and the structure of Yukawa couplings are all encoded geometrically (often topologically) as structures of singularities in the compactification manifold. Although each of the individual, {\it ultra-local} patches giving rise to specific phenomenologically desirable matter representations and specific Yukawa couplings are well understood\footnote{This is true at least in F-theory; in M-theory, a description of the explicit geometries generating Yukawa couplings will be given for the first time in this note. But what we will find in M-theory bears such a striking resemblance to its analogue in F-theory, that it cannot have been unanticipated.} \cite{Atiyah:2001qf,Witten:2001uq,Acharya:2001gy,Katz:1996xe,Berglund:2002hw,Beasley:2008dc,Beasley:2008kw,Heckman:2008ads,Heckman:2008rb,Heckman:2008qa,Bourjaily:2007vw,Bourjaily:2007vx,Bourjaily:2007kv}, all the diverse patches necessary for a complete model have yet to be glued together to form a truly explicit, realistic example.

This problem of building a complete model out of disparate ultra-local regions can be avoided altogether if one starts with a single, sufficiently structured local patch. Take for example the local geometry (described in Section \ref{originofthree}) which generates the \mbox{$\mathbf{27}\,\,\mathbf{27}\,\,\mathbf{27}$} coupling in an $E_6$ grand-unified model in F-theory (or M-theory). Each $\mathbf{27}$ of $E_6$ contains a full generation of matter (including extra singlets and exotic, coloured Higgses\footnote{For background on $E_6$-based grand-unified models, see, e.g., \cite{Hewett:1988xc} and the references therein.}); and it is natural to expect that if the singular structure giving rise to $E_6$ gauge theory were `unfolded' to say, $SU_5$, this single cubic coupling would descend to all of the familiar Yukawa couplings of a traditional $SU_5$-model with three generations (see, e.g., \cite{Bourjaily:2007vw,Bourjaily:2007vx,Bourjaily:2007kv}).  

Before we explain how this can be realized explicitly, it is worth noting that in the setup just described, the fact that there are {\it three generations} of matter interacting locally has nothing at all to do with the global topological data of a compact manifold, such as an Euler characteristic or winding-number\footnote{From a more traditional (heterotic) point of view, it could be said that the local model contains precisely {\it one generation} of matter, which transforms in the `$\mathbf{27}\oplus\mathbf{27}\oplus\mathbf{27}(\oplus\mathbf{1}\oplus\mathbf{1}\oplus\mathbf{1})$' of $E_6$.}. And as we will see, having {\it three} $\mathbf{27}$'s-worth of matter is a {\it generic} feature of any general $\widehat{E_8}$-fibration. This is ultimately nothing more than a reflection of pure group theory, a consequence of the familiar fact that the adjoint of $E_8$ branches into its maximal subgroup $E_6\times SU_3$ according to \vspace{-0.2cm}\begin{equation}\mathbf{248}=(\mathbf{78},\mathbf{1})\oplus(\mathbf{1},\mathbf{8})\oplus(\mathbf{27},\bar{\mathbf{3}})\oplus(\bar{\mathbf{27}},\mathbf{3}).\label{e8_to_e6xsu3}\vspace{-0.2cm}\end{equation}
This could of course be nothing more than deceptive numerology---and as we will see, the number of `generations' one can generate locally in F-theory is too flexible to take (\ref{e8_to_e6xsu3}) too seriously. But M-theory lacks this flexibility, making (\ref{e8_to_e6xsu3}) seem all the more suggestive. 

The local models we describe in this paper are built as fibrations of ALE-spaces. From the example just discussed, we should anticipate that $\widehat{E_8}$-fibrations will be rich enough to incorporate three generations of matter interacting in a phenomenologically interesting way. This is fortunate: the maximality of $\widehat{E_8}$ prevents us from constructing concrete models based on anything more complex, at least in a single local patch. And yet, the degree of complexity needed for any realistic model exceeds that typically achievable in local models based on anything less than $\widehat{E_8}$,\footnote{Refer to our discussion in Section \ref{neede8} for a more thorough assessment.}. But this forced degree of complexity is precisely what we should have anticipated for any model based on an ALE-fibred manifold: the only possible compactification of an ALE-fibre is $K3$---the unique compact Calabi-Yau two-fold---and $K3$ contains two complete copies of the $\widehat{E_8}$ ALE-space (for a review, see e.g., \cite{Aspinwall:1995zi,Aspinwall:1995vk}).

So far we have made no distinction between geometrically engineered models in F-theory and M-theory. Indeed, it turns out that much of the relevant phenomenological data, such as where different matter representations are supported in the manifold and how they are coupled in the superpotential, can be described in a purely algebraic language which is insensitive to whether we are discussing an ALE-fibred $G_2$-manifold for M-theory or an ALE-fibred Calabi-Yau four-fold for F-theory. Because this abstract language is very useful for model building in either framework, we will take care to introduce it in detail in Section \ref{toolbox}. And in \mbox{Section \ref{sectexempli}}, we work through the concrete example of a generic $\widehat{E_7}$-fibration giving rise to $SU_5$ gauge symmetry, which can be realized in both F-theory and M-theory. Those experts familiar with the language used in geometric engineering may prefer to skip \mbox{Section \ref{toolbox}}.

But although we have chosen to describe the geometries of F-theory and M-theory models in similar terms, the two frameworks have many important differences. A summary of the known rules of model building in each framework, emphasizing their differences is given in Section \ref{fmdiffs}. Broadly speaking, models in F-theory have a great deal more flexibility than those in M-theory, leading to many more examples. 

In Section \ref{sigma3andtripleI}, we describe how the triple-intersections of enhanced singularities giving rise to Yukawa couplings in F-theory correlate with the supersymmetric three-cycles which generate Yukawa couplings in M-theory. By doing so, we will provide concrete M-theory analogues for each of the local geometries known and exploited in F-theory phenomenology, filling a gap in the M-theory model-building literature.

In Section \ref{exempliModels} we present several explicit examples of models in F-theory and M-theory that include three generations of matter with realistic phenomenology. Our examples in F-theory are presented somewhat pedagogically, and lead up to two different concrete realizations of the `Diamond Ring' model discussed in Ref.\ \cite{Heckman:2008ads}. This model is quite similar in content to the earlier models described in Ref.\ \cite{Marsano:2008jq}; it includes the possibility of dynamical supersymmetry breaking with gauge-mediation, a dynamical solution to the $\mu/B\mu$-problem, a Peccei-Quinn symmetry with an invisible axion to solve the strong CP-problem, and potentially realistic quark masses. 
The M-theory example presented in \mbox{Section \ref{mtheg}} is based on a high-scale MSSM, and makes use of dynamically-generated effective operators to achieve doublet-triplet splitting and a semi-realistic structure of quark masses. 

Although what we describe in this paper may be merely a lamp-post of concrete examples, it is a lamp-post explicit enough to be exhaustibly analyzed, and yet rich enough to include some of the most realistic string models to date.

\vspace{-0.2cm}
\subsection{A Local Model Builder's Apology}\vspace{-0.1cm}
Interacting, massless chiral matter charged under non-Abelian gauge symmetries are unavoidable ingredients in any realistic model of the Universe. There are roughly three frameworks in which these ingredients are found in string theory: the heterotic string, the type II string on intersecting stacks of D-branes, and geometrically engineered models in F-theory and M-theory. Despite the lack of attention given to geometrically engineered models until recently, they encompass an enormous variety of models, including most of those of the other two frameworks: intersecting D-brane models in type IIa and IIb have geometrically engineered duals in M-theory and F-theory, respectively; and virtually all heterotic models are dual to geometrically engineered models in M-theory\footnote{The equivalence between $E_8\times E_8$ heterotic string theory on $T^3$ and M-theory on $K3$ extends to a fibre-wise duality between heterotic models on $T^3$-fibred Calabi-Yau three-folds and M-theory on $K3$-fibred $G_2$-manifolds. But any Calabi-Yau three-fold that participates in mirror symmetry---and almost all do---can be written as a $T^3$-fibration; therefore almost all heterotic models dualize to M-theory on $K3$-fibred $G_2$-manifolds---models in which gauge theory and chiral matter are geometrically engineered (see, e.g., \cite{Acharya:2004qe}).} \cite{Acharya:2004qe}. 

Perhaps one of the principle reasons why geometrically engineered models have received so little attention in the model-building literature is that almost any physics that results from singularities is inherently {\it localized} in the manifold; and so such models are often best described (or can only be described) in the decompactification limit, $M_{pl}\to\infty$, where the effects of quantum gravity can be ignored. Because of this, local string models are inherently incomplete, and so there may be a danger of confusing a swampland with the landscape---it being very hard to anticipate which local geometries can be ultimately embedded in compact manifolds. And so it may be that the successes of the local models we describe in this paper are destroyed by global obstructions imposed by compactification. Nonetheless, so long as low-scale physics appears insensitive to quantum gravity, in the spirit of \cite{Verlinde:2005jr}, we are encouraged to take the \mbox{$M_{pl}\to\infty$} limit seriously. And doing so can be surprisingly predictive. 

In the decompactification limit, geographically-isolated sectors decouple completely, and this greatly limits both the number of essentially different matter fields that can mutually interact and the number (and types) of gauge-groups under which they can be charged. In particular, local models in F-theory and M-theory cannot arbitrarily support additional gauge symmetries under which the Standard Model fields are charged, or semi-sequestered hidden-sectors with messengers, useful for both gauge-mediated supersymmetry breaking---or explaining dark matter \cite{ArkaniHamed:2008qn}. Indeed, in M-theory the largest locally-achievable gauge groups under which matter can be charged are the rank-seven\footnote{In F-theory, there must be at least two $U_1$-factors (to parameterize the base), and so the largest gauge groups in F-theory with interacting charged matter are the rank-six subgroups of $E_8$. Here, and throughout this paper, we will only include non-Abelian factors when counting rank. The reason for this is that in both F-theory and M-theory, $U_1$-factors are not necessarily associated with gauged-symmetries, but may merely represent (approximate) global symmetries.} subgroups of $E_8$, such as \vspace{-0.3cm}\begin{equation}E_8\supset SU_5\times SU_3\times SU_2\times U_1.\label{subgroup}\vspace{-0.3cm}\end{equation} 
It is tempting to see in this a unified Standard Model with an $SU_3$ family-symmetry and an $SU_2\times U_1$ dark sector. But while local models can indeed accommodate the kinds of theories directly motivated by low-energy physics, they can only barely do so: all of the matter charged under the gauge group (\ref{subgroup}) or its subgroups would be completely decoupled from any other sectors in the theory except through gravity\footnote{This could be useful in M-theory, for example, where supersymmetry breaking could be mediated through gravity mediation from completely sequestered hidden-sectors undergoing gaugino-condensation (see, e.g., \cite{Acharya:2006ia,Acharya:2008zi,Acharya:2007rc}).}. 

Another reason why there have been few phenomenological applications of geometrical engineering until recently is that while string-geometric dualities directly lead to descriptions of the {\it ultra}-local structures that generate particular, individual massless matter representations in F-theory and M-theory, it has not been clear how these structures could be sewn together to form an explicit local geometry including many interacting matter fields. The only known way to explicitly construct a geometry that has multiple, interacting matter fields is to geometrically `unfold' it out of a more singular geometry with more gauge symmetry \cite{Bourjaily:2007kv,Bourjaily:2007vx,Bourjaily:2007vw}. This is the class of models that we discuss in this paper.

Although low-scale intuition can lead to particularly-desirable cartoon `quilts' of ultra-local patches, one should be cautious about using such intuition. This is because phenomenology almost always requires that these patches be sewn together in {\it topologically non-generic} ways in the compactification manifold. Take M-theory for example, where interactions in the superpotential are generated by Euclidean M2-brane instantons wrapping three-cycles that support multiple conical singularities. 
Even allowing for many such three-cycles, in order for a field such as $H^u$ to appear in the superpotential many times, it must lie on many simultaneously-intersecting three-cycles. But three cycles don't generically intersect in a seven-manifold at all\footnote{It is useful to notice that the problem of forcing three-cycles to intersect in a seven-manifold is of the same co-dimensionality as that of forcing lines to intersect in $\mathbb{R}^3$. See Appendix \ref{collininM}.}! Similar comments apply to triple-intersections of complex curves in the two-cycle base of F-theory. But as we will show in Section \ref{sigma3andtripleI}, it turns out that any generic ALE-fibred manifold {\it automatically} possesses these structures. Therefore, if we desire to construct models relevant to phenomenology, then it is reasonable to focus our attention on those models that can be described by such fibrations.

\section[Geometrical Engineering in F-Theory and M-Theory]{Geometrical Engineering in F-Theory and M-Theory:\\ \mbox{A Model Builder's Toolbox}}\label{toolbox}
The models we construct here are based on ALE-\mbox{($K3$-)}fibred compactification manifolds, where the moduli of the ALE-fibres control singularities that vary over the base. The standard description of these fibrations in the geometrical engineering literature is based on the presentation of ALE moduli-spaces given in \mbox{Ref.\ \cite{Katz:1992ab}}, but we are unaware of any comprehensive introduction to its use in model building, or its application to constructing phenomenologically-complete models based on, e.g., $\widehat{E_8}$ (but see \cite{Bourjaily:2007vx}). Therefore, in this section we present a pragmatist's review of this framework---intended for those unfamiliar with it---including a pedagogical example in Section \ref{sectexempli} of a generic $\widehat{E_7}$-fibred model with $SU_5$ gauge symmetry. Those already fluent in this language may prefer to skip to Section \ref{fmdiffs}.
\subsection{The Geometry and Topology of $\widehat{E_n}$-ALE Spaces}\label{sect2}
\begin{table}[b]\begin{center}\hspace{-1.5cm}
\begin{tabular}{rcc@{}c}
\cline{2-4}
&$\begin{array}{c}\text{Positive Roots of~}E_n,\\\text{Labels~of~Two-Cycles~in~}\widehat{E_n}\end{array}$&&$\begin{array}{c}\text{`Area' of Corresponding}\\\text{Two-Cycle in~}\widehat{E_n}(f_1,\ldots,f_n)\end{array}$\\
\cline{2-4}& $e_i-e_{j>i}$&$\implies$&$f_i-f_{j>i}$\\
&$-e_0+e_i+e_j+e_k$&$\implies$& $f_i+f_j+f_k$\\
\footnotesize{$n\geq 6$}&$-2e_0+\sum_{j=1}^6e_{i_j}$ &$\implies$& $\sum_{j=1}^6f_{i_j}$\\
\footnotesize{$n=8$}&$-3e_0+e_{i}+\sum_{j=1}^8e_{j}$&$\implies$& $f_i+\sum_{j=1}^8f_j$\\
\cline{2-4}
\end{tabular}\caption{\label{En_roots_and_areas}The roots of $E_n$---written as vectors in $\mathbb{R}^{n,1}$ having squared-norm of 2---which are in one-to-one correspondence with two-cycles in $\widehat{E_n}$ whose areas are fixed by the moduli $(f_1,f_2,\ldots,f_n)$. This is adapted from Table 4 of Ref.\ \cite{Katz:1992ab} and reviewed in Appendix \ref{ALEApp}. }\end{center}\end{table}
The Asymptotically Locally Euclidean (ALE) space $\widehat{E_n}$ is the non-compact, hyper-K\"{a}hler Calabi-Yau two-fold which is the desingularization of the orbifold $\mathbb{C}^2/\Gamma_{E_n}$ (these co-dimension four Kleinian orbifolds are reviewed in Appendix \ref{ALEApp}). $\widehat{E_n}$ is so named because it possesses an entire $E_n$ root-lattice of supersymmetric two-cycles; by this we mean that a basis of two-cycles in $\widehat{E_n}$ can be chosen so that their intersection matrix is simply minus the Cartan matrix of the $E_n$ Lie algebra. In this basis, two-cycles in $\widehat{E_n}$ can be labelled by roots of $E_n$, and two two-cycles intersect iff their corresponding root-labels have a non-vanishing inner product. The positive roots of $E_n$---which name (the `positive' half of) the supersymmetric two-cycles in $\widehat{E_n}$---are listed on the left-hand side of \mbox{Table \ref{En_roots_and_areas}}.

Consider for example the space $\widehat{E_7}$. Using \mbox{Table \ref{En_roots_and_areas}} it is easy to enumerate the roots in the $E_7$-lattice: there are \mbox{${\scriptsize\left(\begin{array}{c}7\\2\end{array}\right)}+{\scriptsize\left(\begin{array}{c}7\\3\end{array}\right)}+{\scriptsize\left(\begin{array}{c}7\\6\end{array}\right)}=63$} positive roots, making 126 roots in total. If all were shrunk to zero size---by setting all the moduli to zero---then F-theory or M-theory compactified on such a space would give rise to $E_7$ gauge theory. In M-theory, for example, the worldlines of massless membranes wrapping these 126 shrunk two-cycles would combine with the seven vector-fields from Kaluza-Klein reduction of the supergravity three-form on the $b_2(\widehat{E_7})=7$ basis two-cycles to fill out the $\mathbf{133}$ of $E_7$, its adjoint representation (see, e.g., \cite{Acharya:2004qe}).

Naming the two-cycles in $\widehat{E_7}$ according to \mbox{Table \ref{En_roots_and_areas}} has several important advantages. Because the intersection-number of a pair of two-cycles is given by minus the (mostly-plus) Minkowski inner product of their root-lattice labels, it is relatively easy to visualize how they are `physically arranged' in the $\widehat{E_7}$-space. Let's see how this works. A choice of {\it simple} roots of $E_7$ constitutes a choice of basis for the second homology class of $\widehat{E_7}$; if we choose this basis to be the two-cycles labelled by,
\begin{equation}
\left\{\begin{array}{l}\hspace{3.65cm}(-e_0+e_5+e_6+e_7)\\(e_1-e_2)\,(e_2-e_3)\,(e_3-e_4)\,(e_4-e_5)\,(e_5-e_6)\,(e_6-e_7)\end{array}\right\},
\label{e7dynk}\end{equation}
we find that they are {\it physically arranged} according the the $E_7$ Dynkin diagram\footnote{The Dynkin diagram is generated by drawing a node for each basis two-cycle, and connecting the nodes of intersecting cycles. For example, $(-e_0+e_5+e_6+e_7)\cdot(e_4-e_5)=-1$ while $(-e_0+e_5+e_6+e_7)$ is orthogonal to all of the other simple roots in (\ref{e7dynk}); therefore, its node---on the top of (\ref{s2dynk})---is connected to only that of $(e_4-e_5)$---the third node from the right.}: \vspace{-0.2cm}\begin{equation}\mbox{$\hspace{-3.4cm}\bigcirc\hspace{-1pt}\raisebox{2.pt}{\rule{37pt}{1.5pt}}\hspace{-3.35pt}\bigcirc\raisebox{2.pt}{\hspace{-3.35pt}\rule{37pt}{1.5pt}}\hspace{-3.35pt}\bigcirc\raisebox{2.pt}{\hspace{-3.35pt}\rule{37pt}{1.5pt}}\hspace{-3.35pt}\bigcirc\raisebox{2.pt}{\hspace{-3.35pt}\rule{37pt}{1.5pt}}\hspace{-3.35pt}\bigcirc\raisebox{2.pt}{\hspace{-3.35pt}\rule{37pt}{1.5pt}}\hspace{-3.35pt}\bigcirc\,\,.\raisebox{8.35pt}{\hspace{-3.94cm}\rule{1.5pt}{18pt}\raisebox{20.35pt}{\hspace{-6.7pt}$\bigcirc$}}$}\label{s2dynk}\vspace{-0.2cm}\end{equation}
This allows for a powerful dictionary between geometry and the resulting physics: a Dynkin diagram-worth of shrunk two-cycles gives rise to the corresponding gauge symmetry; and as we will see later, two-cycles {\it that shrink at localized places on the base} fill out weight-lattices of corresponding matter representations that live where the two-cycles vanish. 

\subsection{Exempli Gratia: $E_7\to SU_5\times U_1^a\times U_1^b\times U_1^c$}\label{sectexempli}

It is worthwhile to work through at least one complete example pedagogically. Once this is done, it should be straight-forward to see how such an analysis can be automated for models based on any arbitrary fibration of $\widehat{E_n}$-spaces, allowing similar calculations to be done on a computer\footnote{We have written Mathematica code which automates these calculations, and will gladly share it with those who are interested.}.

The example we wish to analyze here is the geometrically engineered manifold specified by the fibration \mbox{$\widehat{E_7}(a,b,c,0,0,0,0)$}, where $a,b,$ and $c$ are maps from the base space $W$---either $\mathbb{R}^3$ or $\mathbb{C}^2$---into the first three of the seven moduli\footnote{The question of whether we consider the moduli $(f_1,\ldots,f_7)$ to be the full hyper-K\"{a}hler moduli of $\widehat{E_7}$ or complex structure moduli of an equivalent hypersurface in $\mathbb{C}^3$ will be discussed below.} of $\widehat{E_7}$ in Katz and Morrison's basis \cite{Katz:1992ab}; the other four moduli are fixed at zero everywhere over the base. What we mean is, we construct the total manifold by gluing over each point $p\in W$ the fibre \mbox{$\widehat{E_7}(a(t),b(t),c(t),0,0,0,0)|_{t=p}$}. This is reviewed in \mbox{Appendix \ref{ALEApp}}.
 
Let us first study the structure of a generic fibre \mbox{$\widehat{E_7}(a,b,c,0,0,0,0)$} over a fixed point of the base. Having generic, non-vanishing moduli $a,b,$ \mbox{and $c$} will obviously blow up some of the 126 two-cycles, leaving us with less than $E_7$ gauge symmetry. Because four of the moduli vanish, we see at once that there will be (at least) an $E_4(\equiv SU_5)$-lattice-worth of shrunk cycles---the ten `positive' ones of which are:\vspace{-0.2cm}
\begin{equation}
{\normalsize\begin{array}{c}
\text{\raisebox{10pt}{$\displaystyle \int_\Sigma\omega=0$}}\\
\hline
(-e_0+e_4+e_5+e_6)\\
(e_4-e_7)\quad(-e_0+e_4+e_5+e_7)\\
(e_5-e_7)\quad(e_4-e_6)\quad(-e_0+e_4+e_6+e_7)\\
(e_6-e_7)\quad(e_5-e_6)\quad(e_4-e_5)\quad(-e_0+e_5+e_6+e_7)\vspace{-0.1cm}\\
\|\hspace{47pt}\|\hspace{49pt}\Arrowvert\hspace{76pt}\|\hspace{30pt}\vspace{-0.1cm}\\
\alpha_1\hspace{43pt}\alpha_2\hspace{43pt}\alpha_3\hspace{70pt}\alpha_4\hspace{30pt}\vspace{-0.1cm}\\
\bigcirc\hspace{-1pt}\raisebox{2.pt}{\rule{43pt}{1.5pt}}\hspace{-3.35pt}\bigcirc\raisebox{2.pt}{\hspace{-3.5pt}\rule{43pt}{1.5pt}}\hspace{-3.25pt}\bigcirc\raisebox{2.pt}{\hspace{-3.45pt}\rule{70pt}{1.5pt}}\hspace{-1.05pt}\bigcirc\hspace{30pt}\end{array}}\label{eg_su5_lattice}
\vspace{-0.3cm}\end{equation}
Here, we have used $\omega$ to denote the entire $SU_2$-triplet of K\"{a}hler forms on $\widehat{E_7}$---which exist for any hyper-K\"{a}hler manifold---and so `$\int_{\Sigma}\omega$' represents a {\it three-component} generalization of the `{\it area}' of a two-cycle $\Sigma$ (see, e.g., \cite{Hitchin:1994zr}). In (\ref{eg_su5_lattice}), we have listed the two-cycles with vanishing `areas' according to (the positive-half of) the weight-diagram for the $SU_5$-adjoint to make the root-lattice more apparent. When writing it this way, the bottom-row is the set of {\it simple} roots of $SU_5$, with higher rows obtained by adding positive multiples of the simple roots to each row successively\footnote{For background on Lie algebras and their representations, see, e.g., \cite{Fuchs:1997jv,Slansky:1981yr,FultonRepTheory}.}. We've chosen to label the simple roots of $SU_5$ as $\alpha_{1,\ldots,4}$, and we will sometimes refer to a root $r$ by its {\it Dynkin label}, which is simply the four-tuple $(r\cdot\alpha_1\,\,\,\,r\cdot\alpha_2\,\,\,\,r\cdot\alpha_3\,\,\,\,r\cdot\alpha_4)$. For example, the root $(e_5-e_7)=\alpha_1+\alpha_2$ has Dynkin label $(1\,1\,\text{-}1\,0)=(2\,\text{-}1\,0\,0)+(\text{-}1\,2\,\text{-}1\,0)$.

All the other two-cycles in $\widehat{E_7}$ have `areas' that are generally non-vanishing functions of $a,b,$ \mbox{and $c$}. Listing them below according to area (calculated using \mbox{Table \ref{En_roots_and_areas}}), we see that they fall nicely into the weight-diagrams of familiar representations of $SU_5$. The 53 non-vanishing, positive two-cycles in $\widehat{E_7}(a,b,c,0,0,0,0)$ are\footnote{Recall that the weight-lattice of a representation is generated by starting with the highest-weight and successively subtracting each simple root $\alpha_i$ that intersects it negatively.}: \vspace{-0.2cm}

\begin{itemize}
\item those whose highest-weight's Dynkin label\footnote{Here, two-cycles are arranged according to ordinary weight-diagrams (`spindles'), with the highest-weight listed on the top. For example, the highest-weight two-cycle of those with area $a$ has a root-label \mbox{$r_a=(-e_0+e_1+e_4+e_5)$}, and $(r_a\cdot\alpha_1\,\,\,\,r_a\cdot\alpha_2\,\,\,\,r_a\cdot\alpha_3\,\,\,\,r_a\cdot\alpha_4)=(0100)$.} is $(0100)$, which form $\mathbf{10}$'s of $SU_5$:
\begin{equation*}\hspace{-1.85cm}{\scriptsize\begin{array}{cccccccc}
\multicolumn{2}{c}{\text{\raisebox{8pt}{\scalebox{1.35}{$\displaystyle\int_{\Sigma}\omega=a$}}}}&&\multicolumn{2}{c}{\text{\raisebox{8pt}{\scalebox{1.35}{$\displaystyle\int_{\Sigma}\omega=b$}}}}&&\multicolumn{2}{c}{\text{\raisebox{8pt}{\scalebox{1.35}{$\displaystyle\int_{\Sigma}\omega=c$}}}}\\\cline{1-2}\cline{4-5}\cline{7-8}\multicolumn{2}{c}{(-e_0+e_1+e_4+e_5)}&&\multicolumn{2}{c}{(-e_0+e_2+e_4+e_5)}&&\multicolumn{2}{c}{(-e_0+e_3+e_4+e_5)}\\
\multicolumn{2}{c}{(-e_0+e_1+e_4+e_6)}&&\multicolumn{2}{c}{(-e_0+e_2+e_4+e_6)}&&\multicolumn{2}{c}{(-e_0+e_3+e_4+e_6)}\\
(-e_0+e_1+e_5+e_6)&(-e_0+e_1+e_4+e_7)&&(-e_0+e_2+e_5+e_6)&(-e_0+e_2+e_4+e_7)&&(-e_0+e_3+e_5+e_6)&(-e_0+e_3+e_4+e_7)\\
(e_1-e_7)&(-e_0+e_1+e_5+e_7)&&(e_2-e_7)&(-e_0+e_2+e_5+e_7)&&(e_3-e_7)&(-e_0+e_3+e_5+e_7)\\
(e_1-e_6)&(-e_0+e_1+e_6+e_7)&&(e_2-e_6)&(-e_0+e_2+e_6+e_7)&&(e_3-e_6)&(-e_0+e_3+e_6+e_7)\\
\multicolumn{2}{c}{(e_1-e_5)}&&\multicolumn{2}{c}{(e_2-e_5)}&&\multicolumn{2}{c}{(e_3-e_5)}\\
\multicolumn{2}{c}{(e_1-e_4)}&&\multicolumn{2}{c}{(e_2-e_4)}&&\multicolumn{2}{c}{(e_3-e_4)}
\end{array}}\end{equation*}
\item those with highest-weight label $(0001)$, corresponding to $\bar{\mathbf{5}}$'s of $SU_5$:
\vspace{-0.2cm}\begin{equation*}\hspace{-1.9cm}{\footnotesize\begin{array}{ccccc}
\text{\raisebox{8pt}{\scalebox{1.15}{$\displaystyle\int_{\Sigma}\omega=a+b$}}}&&\text{\raisebox{8pt}{\scalebox{1.15}{$\displaystyle\int_{\Sigma}\omega=a+c$}}}&&\text{\raisebox{8pt}{\scalebox{1.15}{$\displaystyle\int_{\Sigma}\omega=b+c$}}}\\
\cline{1-1}\cline{3-3}\cline{5-5}(-2e_0+e_1+e_2+e_4+e_5+e_6+e_7)&&(-2e_0+e_1+e_3+e_4+e_5+e_6+e_7)&&(-2e_0+e_2+e_3+e_4+e_5+e_6+e_7)\\
(-e_0+e_1+e_2+e_4)&&(-e_0+e_1+e_3+e_4)&&(-e_0+e_2+e_3+e_4)\\
(-e_0+e_1+e_2+e_5)&&(-e_0+e_1+e_3+e_5)&&(-e_0+e_2+e_3+e_5)\\
(-e_0+e_1+e_2+e_6)&&(-e_0+e_1+e_3+e_6)&&(-e_0+e_2+e_3+e_6)\\
(-e_0+e_1+e_2+e_7)&&(-e_0+e_1+e_3+e_7)&&(-e_0+e_2+e_3+e_7)
\end{array}}\vspace{-0.4cm}\end{equation*}
\item one set of cycles with highest-weight label $(1000)$, a $\mathbf{5}$ of $SU_5$:
\vspace{-0.2cm}\begin{equation*}{\footnotesize\begin{array}{c}
\text{\raisebox{8pt}{\scalebox{1.15}{$\displaystyle\int_{\Sigma}\omega=a+b+c$}}}\\
\hline
(-2e_0+e_1+e_2+e_3+e_4+e_5+e_6)\\
(-2e_0+e_1+e_2+e_3+e_4+e_5+e_7)\\
(-2e_0+e_1+e_2+e_3+e_4+e_6+e_7)\\
(-2e_0+e_1+e_2+e_3+e_5+e_6+e_7)\\
(-e_0+e_1+e_2+e_3)
\end{array}}
\vspace{-0.4cm}\end{equation*}
\item and three singlets of $SU_5$, each with Dynkin label $(0000)$:
\vspace{-0.2cm}\begin{equation*}\begin{array}{ccccc}
\text{\raisebox{7pt}{\scalebox{1.}{$\displaystyle\int_{\Sigma}\omega=a-b$}}}&&\text{\raisebox{7pt}{\scalebox{1.}{$\displaystyle\int_{\Sigma}\omega=a-c$}}}&&\text{\raisebox{7pt}{\scalebox{1.}{$\displaystyle\int_{\Sigma}\omega=b-c$}}}\\
\cline{1-1}\cline{3-3}\cline{5-5}(e_1-e_2)&&(e_1-e_3)&&(e_2-e_3)
\end{array}\vspace{-0.2cm}
\end{equation*}
\end{itemize}

In addition to the 53 non-vanishing two-cycles listed above, there are also those corresponding to the negative roots. They have opposite areas\footnote{That is, their areas are oppositely-oriented with respect to the triplet of K\"ahler forms.}, and transform in representations conjugate to the ones listed above. For example, the `negative' two-cycle $(-e_1+e_4)$ has area $-f_1+f_4=-a$  and has Dynkin label $(0010)$. This means that it represents the highest-weight of an entire $\bar{\mathbf{10}}$-worth of two-cycles that all have area $-a$. Therefore, considering both the positive and negative roots, non-vanishing two-cycles in the geometry come in `hypermultiplets.' So why go through the trouble of distinguishing a $\mathbf{10}$ of two-cycles with area $a$ from a $\bar{\mathbf{10}}$ with area $-a$? The answer is that the sign of a two-cycle's area correlates with its conjugation; this will be seen to essentially keep track of $U_1$-charges, and so is crucial in determining which operators can appear in the superpotential.

Although there are $53\times2$ non-vanishing two-cycles in general, only three of them are independent\footnote{This is because all of $\widehat{E_7}$ has only seven independent two-cycles, of which four form the $SU_5$ sub-lattice of shrunk two-cycles (\ref{eg_su5_lattice}). }, and we are free to choose any convenient combination as a basis. Because $U_1$-symmetries arise from generally non-vanishing two-cycles, this amounts to a choice of basis for a $U_1\times U_1\times U_1$ symmetry in the theory. One natural choice for these corresponds to the set of two-cycles with areas $a,b,$ and $c$---those of the three $\mathbf{10}$'s of $SU_5$. Making this choice, the generically non-vanishing two-cycles in the space $\widehat{E_7}(a,b,c,0,0,0,0)$ are listed in \mbox{Table \ref{exempli_table}}.
\newpage
\vspace{-0.0cm}\begin{table*}[t]\begin{center}\caption{Generically non-vanishing two-cycles in the fibration $\widehat{E_7}(a,b,c,0,0,0,0)$.\label{exempli_table}}\vspace{0.4cm}
\begin{tabular}{|r@{\hspace{0.5cm}}llll|c|}
\hline
&$\!\!SU_5\,\times\,\!\!\!\!$&$\!U_1^a\,\times\!\!\!\!$&$\!U_1^b\,\times\!\!\!\!$&$\!U_1^c$&Area $(\int_{\Sigma}\omega$)\\\hline
\crowF{T_1}{10}{1}{0}{0}{a}
\crowF{T_2}{10}{0}{1}{0}{b}
\crowF{T_3}{10}{0}{0}{1}{c}
\crowF{F_1}{\bar{5}}{1}{1}{0}{a+b}
\crowF{F_2}{\bar{5}}{1}{0}{1}{a+c}
\crowF{F_3}{\bar{5}}{0}{1}{1}{b+c}
\crowF{F_4}{5}{1}{1}{1}{a+b+c}
\crowF{S_1}{1}{1}{-1}{0}{a-b}
\crowF{S_2}{1}{1}{0}{-1}{a-c}
\crowF{S_3}{1}{0}{1}{-1}{b-c}
\hline
\end{tabular}\end{center}\end{table*}
~\\\vspace{-1.5cm}

It is worth noting that the content of \mbox{Table \ref{exempli_table}} is nothing more than pure representation theory: it merely reflects the branching of the adjoint of $E_7$ into its subgroup \mbox{$SU_5\times U_1^a\times U_1^b\times U_1^c$},
\vspace{-0.3cm}\begin{equation*}\begin{split}
\hspace{-1cm}\mathbf{133}=&\phantom{{}\oplus{}}\mathbf{24}_{0,0,0}\oplus\mathbf{1}_{0,0,0}+\mathbf{1}_{0,0,0}\oplus\mathbf{1}_{0,0,0}\\
&\oplus\mathbf{10}_{1,0,0}\hspace{2.85pt}\oplus\mathbf{10}_{0,1,0}\hspace{2.85pt}\oplus\mathbf{10}_{0,0,1}\hspace{2.85pt}\oplus\bar{\mathbf{5}}_{1,1,0}\hspace{5.7pt}\oplus\bar{\mathbf{5}}_{1,0,1}\hspace{5.7pt}\oplus\bar{\mathbf{5}}_{0,1,1}\hspace{5.7pt}\oplus\mathbf{5}_{1,1,1}\hspace{8.55pt}\oplus\mathbf{1}_{1,\text{-}1,0}\oplus\mathbf{1}_{1,0,\text{-}1}\oplus\mathbf{1}_{0,1,\text{-}1}\\
&\oplus\bar{\mathbf{10}}_{\text{-}1,0,0}\oplus\bar{\mathbf{10}}_{0,\text{-}1,0}\oplus\bar{\mathbf{10}}_{0,0,\text{-}1}\oplus\mathbf{5}_{\text{-}1,\text{-}1,0}\oplus\mathbf{5}_{\text{-}1,0,\text{-}1}\oplus\mathbf{5}_{0,\text{-}1,\text{-}1}\oplus\bar{\mathbf{5}}_{\text{-}1,\text{-}1,\text{-}1}\oplus\mathbf{1}_{\text{-}1,1,0}\oplus\mathbf{1}_{\text{-}1,0,1}\oplus\mathbf{1}_{0,\text{-}1,1}.\\
\end{split}\end{equation*}
This should not be surprising: all our work above has just been a geometric version of a completely standard root-lattice calculation. But although these calculations are quite straight-forward (if tedious) exercises in representation theory, the detailed branching displayed in \mbox{Table \ref{exempli_table}}, which includes all of the $U_1$-charges, goes beyond the detail that would be found in familiar physics resources, such as Slanksy's review \cite{Slansky:1981yr}. Therefore, analyses such as the example above can be useful for model building in practice, where $U_1$-charges can be very important. 

Now that we understand the entire lattice of two-cycles in the fibre over a particular point, let us turn to a description of the complete fibration \mbox{$\widehat{E_7}(a,b,c,0,0,0,0)$}. But in order to describe the total space, we must differentiate between the bases used in F-theory and M-theory. When building an $\widehat{E_7}$-fibred $G_2$-manifold for M-theory, it is important to retain the full hyper-K\"{a}hler structure of $\widehat{E_7}$, and so we take all the moduli to be elements of $\mathbb{R}^3$ (or sometimes, more elegantly, the imaginary quaternions), constructing geometries as hyper-K\"{a}hler quotients \cite{Acharya:2001gy}. In F-theory, however, we start by fixing a complex structure of $\widehat{E_7}$, and use complex structure deformations of its corresponding Kleinian hypersurface to construct our compactification manifold as a fibration over $\mathbb{C}^2$ \cite{Katz:1996xe}. The connection between these two approaches is reviewed in \mbox{Appendix \ref{ALEApp}}. In both cases there is a notion of holomorphicity that must be obeyed by the maps $a,b,c$---whether as ordinary complex maps or as maps of the quaternions.

Now, so long as $a,b,$ and $c$ are non-constant maps, there will be places over the base where each of the sets of two-cycles in \mbox{Table \ref{exempli_table}} have vanishing area. Wherever this happens along the base, there will be matter fields transforming in the representation following from the weight-lattice of shrinking two-cycles\footnote{This can be understood in type IIa and F-theory by solving the Dirac equation directly \cite{Katz:1996xe}, or in M-theory by duality with the heterotic string \cite{Acharya:2001gy}.}. For example, letting $t$ denote a coordinate on the base, there will be a $\mathbf{10}$ (and/or a $\bar{\mathbf{10}}$) of matter resulting from the $\mathbf{10}$ (and $\bar{\mathbf{10}}$) of two-cycles that shrink wherever $a(t)=0$. In F-theory, where the base is a complex two-fold, matter is localized along complex curves, while in M-theory, where the base is a real three-cycle, matter is isolated at points---where in both cases the degree of the function $a(t)$ determines the multiplicity of matter along the base. Partly to avoid this multiplicity and the resulting complexity of higher-degree maps, we will limit our attention in this paper to only linear functions of the coordinates of the base---linear, but not strictly proportional. 

\begin{figure}[!t]\begin{center}\caption{The lines of various enhanced singularities over the base of the fibration $\widehat{E_7}(a(t),b(t),c(t),0,0,0,0)$, where $t\in\mathbb{C}^2$. Every fibre over the plane has at least an $SU_5$-singularity. Thick, solid blue lines correspond to the locations where lattices of $\mathbf{10}$'s (or $\bar{\mathbf{10}}$'s) shrink to zero size; widely-dashed red lines are $\mathbf{5}$ or $\bar{\mathbf{5}}$'s, and thin, finely-dashed  black lines are singlets (which generally lie off the plane). The green line, or `slice,' defines the base space of an analogous model in M-theory, with dots along it indicating the locations of conical singularities supporting matter representations. The parenthetical section along the M-theory slice refers to Figure \ref{three_cycle}.\label{e7_exempli}}\vspace{0.5cm}\mbox{\hspace{-0cm}\includegraphics[scale=1.65,angle=0]{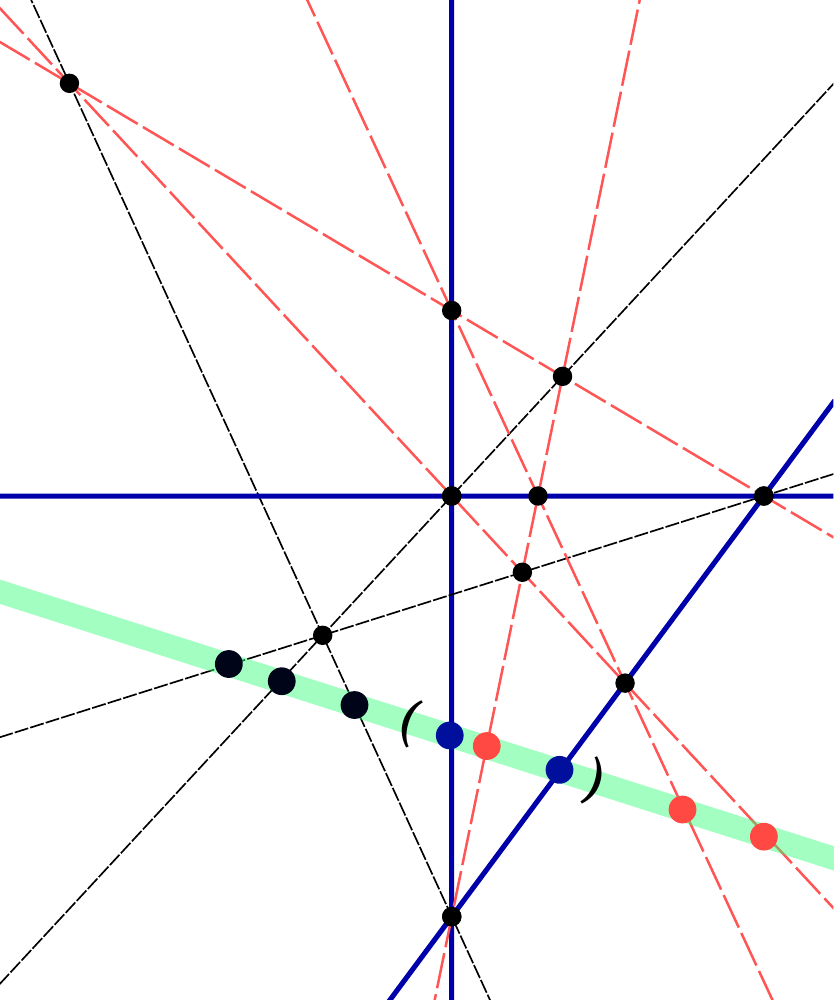}}
\end{center}\end{figure}

\mbox{Figure \ref{e7_exempli}} is an illustration of the fibration $\widehat{E_7}(a,b,c,0,0,0,0)$ for randomly-chosen (linear) maps $a,b,$ and $c$. The figure represents a two-parameter base $W$ over which every fibre has at least an $SU_5$-singularity. But this $SU_5$-singularity is enhanced along the various `matter-curves' such as $a(t)=0$, where it becomes of type $SO_{10}$. The lines in \mbox{Figure \ref{e7_exempli}} represent each of the various matter-curves according to \mbox{Table \ref{exempli_table}}, and are coloured according to representation: thick, solid blue lines being $\mathbf{10}$'s and $\bar{\mathbf{10}}$'s, widely-dashed red lines being $\mathbf{5}$'s and $\bar{\mathbf{5}}$'s, and finely-dashed black lines being $\mathbf{1}$'s; triple-intersections of matter curves indicate possible Yukawa couplings and are labelled by black dots.   \mbox{Figure \ref{e7_exempli}} represents a random point in the moduli-space of an F-theory model based on this fibration---the parameters in the maps $a,b,$ and $c$ being complex structure moduli of the total fibration.

Because the base space in M-theory is described by a single coordinate-parameter, a real three-vector $t$, while the base in F-theory is described by two (complex) coordinate-parameters, the base in M-theory can be imagined as a one-parameter restriction of that of F-theory---as a line, or `slice,' through the F-theory plane. One such slice is indicated by an especially thick green line in \mbox{Figure \ref{e7_exempli}}, and the large dots along it indicate the locations of conical singularities in the M-theory base\footnote{One aspect of the M-theory fibration that is lost in this method of illustration is that the conical singularities in M-theory are not generally collinear---although some are: see Appendix \ref{collininM}.}. Notice that any generic M-theory `slice' through the F-theory plane will intersect each of the matter-curves once. This means that in general, both the F-theory and M-theory models based on a given fibration will have the same matter content---up to the multiplicities allowed in F-theory (see \mbox{Section \ref{fmdiffs2}}). This need not be the case, however: one can always choose a non-generic slice for M-theory that is parallel to one (or more) of the matter-curves, thereby forcing the corresponding conical singularity out of the local geometry. This freedom may be critical for getting rid of unwanted light degrees of freedom in M-theory, and we will use this in \mbox{Section \ref{mtheg}.}

\newpage~

\section{F-Theory vs. M-Theory: Weighing the Options}\label{fmdiffs}\vspace{-0.2cm}

\subsection[Yukawa Couplings in F-Theory and M-Theory]{Yukawa Couplings in F-Theory and M-Theory}\label{sigma3andtripleI}
In F-theory, triple-intersections of matter-curves can generate cubic operators in the superpotential\footnote{The operators may involve massive fields irrelevant to low-scale physics; so far we have not made any distinction among the chiral components of the `hypermultiplets' of shrinking two-cycles in the geometry.}, the coefficients of which come from a triple-overlap integral of wavefunctions living along each of the matter-curves \cite{Beasley:2008dc,Beasley:2008kw}. Importantly, these coefficients are not obviously suppressed. In M-theory, however, terms in the superpotential arise from Euclidean M2-brane instantons wrapping supersymmetric three-cycles in the manifold, and therefore have coefficients that are typically highly suppressed by the instanton action, which is exponential in the volume of the three-cycle.

One thing to notice about Figure \ref{e7_exempli}---the illustration of an example $\widehat{E_7}$-fibration in F-theory---is the ubiquity of multiply-intersecting matter-curves. Despite the na\"{i}ve geometric non-generiticity of triple-intersections, they turn out to be {\it ubiquitous} features of any model constructed via ALE-fibrations. We will show this presently, and follow by demonstrating that such triple-intersections are in correspondence with supersymmetric three-cycles in M-theory that support multiple conical singularities.

The ubiquity of multiple-intersections follows from the fact that in ALE-fibrations, {\it every} intersection between two matter-curves involves at least a third\footnote{In \mbox{Figure \ref{e7_exempli}}, there may appear to be a few double-intersections. Closer inspection will show that these always involve singlets (thinly-dashed black lines); along these matter-curves, the singularity of the fibre is of type $SU_5\times SU_2$, with two singularities separated by a non-vanishing two-cycle. This means that the black curves do not lie in the $SU_5$-plane, and so the apparent double intersections are spurious.}. An easy way to prove this uses representation theory, and such a proof is given in \mbox{Appendix \ref{TIeqGIO}}. Roughly speaking, if two independent two-cycles of the fibre, say $\Sigma_a$ and $\Sigma_b$, shrink simultaneously along the base, then there must also be a distinct, third set of shrinking two-cycles (in the homology class $[\Sigma_a]\pm[\Sigma_b]$).

The converse, however, that every gauge-invariant cubic operator involving matter fields in an ALE-fibred manifold in F-theory will have a corresponding triple-intersection, is easier to prove. As we saw in \mbox{Section \ref{sectexempli}}, $U_1$-charges dictate geography in the manifold. Suppose that the fields $A,\,B,$ and $C$ live at the solutions to $f_{A,B,C}=0$, respectively; if the operator $A\,B\,C$ is gauge invariant, their $U_1$-charges must add up to zero, implying that $f_A+f_B+f_C=0$. Because there will generically be a place on the base where $f_A=f_B=0$---each being a function of two parameters in F-theory---we see that at this place $f_C$ must vanish also. For example, consider operator $T_1\,\bar{F_1}\,T_2$ from the fields in \mbox{Table \ref{exempli_table}}. $T_1,\,T_2$ and $\bar{F_1}$ live along the curves $a=0,\,b=0,$ and $-a-b=0$, respectively. Obviously, when $a=b=0$, $-a-b=0$. 

\begin{figure}[t]\begin{center}\caption{The geometry in M-theory described by the section of the thick, solid green slice in \mbox{Figure \ref{e7_exempli}} enclosed in parentheses. The purple line through the fibres represents the location of the shrunk $SU_5$-lattice of two-cycles in each fibre. The blue dots on either side are the conical singularities giving rise to massless $\mathbf{10}$'s of $SU_5$ and the singularity indicated by the red dot between the two gives rise to a massless $\mathbf{5}$. Notice the foliated three-cycle that supports all three singularities. \label{three_cycle}}\mbox{\hspace{-0.0cm}\includegraphics[width=14cm,height=5.25cm]{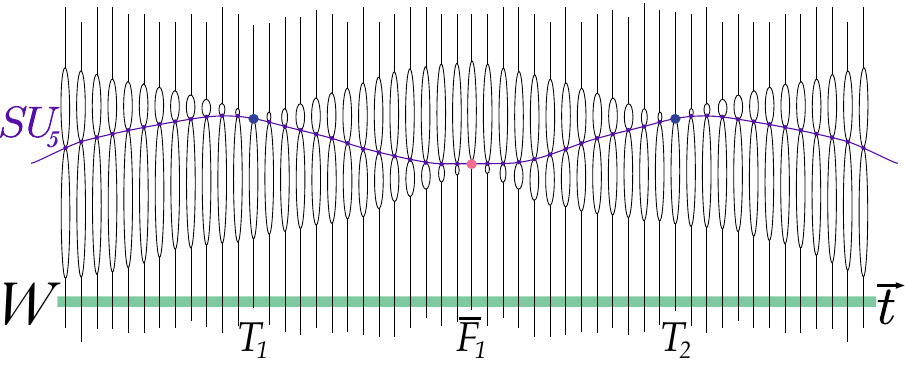}}
\end{center}\vspace{-1.5cm}\end{figure}

Let us now see that this translates into the existence of a supersymmetric three-cycle in an M-theory version of the fibration such as the slice illustrated in \mbox{Figure \ref{e7_exempli}} (the thick, solid green line through the plane). The fields $T_1,\,\bar{F_1},$ and $T_2$ are those enclosed in parentheses along the slice. As we will now show, the geometry of the fibration along the base will always include a supersymmetric three-cycle supporting the three conical singularities as illustrated in \mbox{Figure \ref{three_cycle}}. A simple argument (given in \mbox{Appendix \ref{collininM}}) shows that whenever a cubic operator is gauge-invariant, the conical singularities supporting the three fields are collinear along the three-dimensional base; the supersymmetric three-cycle supporting the singularities is foliated by two-cycles in the fibres over points along the the line that connects them. For example, notice that one of the two-cycles composing $T_1$ is \mbox{$[(-e_0+e_1+e_4+e_5)]\equiv[\Sigma_a]$} and that one of those composing $\bar{F_1}$ is \mbox{$[(e_0-e_1-e_2-e_5)]\equiv[\Sigma_{-a-b}]$}. And, because \mbox{$[\Sigma_{a}]+[\Sigma_{-a-b}]=-[(e_2-e_4)]\equiv-[\Sigma_{b}]$}, which is one of the two-cycles that shrinks within $T_2$, we see that there is foliated three-cycle supporting the three fields.

Because we know the areas of two-cycles in each ALE-fibre with respect to the ALE-hyper-K\"{a}hler metric, insofar as the complete $G_2$-metric can be approximated as a product of that of the base and that of the fibre\footnote{Depending on the size of the three-cycle relative to the compactification scale of the base---which we typically envision as $S^3$ in M-theory---this may or may not be a reasonable assumption. Because the size of the base sets the strength of the gauge theory coupling---$\mathrm{Vol}(S^3)\sim g_{YM}^{-2}$ in Planck units---while the local three-cycle volumes set the sizes of Yukawa couplings relevant to phenomenology, knowing the relative scales of coefficients in the superpotential should allow us to asses the validity of assuming the base to be essentially flat over the fibration. }, it is very easy to integrate the areas to determine the volume of the three-cycle that is foliated from $T_1$ to $T_2$:\vspace{-0.3cm}\begin{equation}\int\limits_{\Sigma_3(T_1\,\bar{F_1}\,T_2)}\!\!\!\!\!\!d\mathrm{Vol}=\left\{\int_{t_{T_1}}^{t_{\bar{F_1}}}a(t)dt+\int_{t_{\bar{F_1}}}^{t_{T_2}}(-b(t))dt\right\}=\tfrac{1}{2}\big|a(t_{\bar{F_1}})\big|\big|(t_{T_2}-t_{T_1})\big|.\label{explicit_yukawa_in_mtheory}\vspace{-0.3cm}\end{equation}
\noindent If $a$ and $b$ are given by $a(t)=\alpha t-a_0~\mathrm{and}~b(t)=\beta t-b_0,$ then this becomes\vspace{-0.2cm}\begin{equation}\int\limits_{\Sigma_3(T_1\,\bar{F_1}\,T_2)}\!\!\!\!\!\!d\mathrm{Vol}=\frac{\left|\alpha b_0-\beta a_0\right|^2}{2\alpha\beta(\alpha+\beta)}.\label{explicit_yukwa_in_mtheory_eg}\vspace{-0.2cm}\end{equation}

This discussion readily generalizes to arbitrary higher-point operators, which can be seen as the M-theory lifts of world-sheet disc instantons in type IIa string theory (see, e.g., \cite{Cvetic:2003ch}).  
Let us now show that there exists a supersymmetric three-cycle supporting the set of conical singularities $\{X_i\}$ whenever the operator $\prod_iX_i$ is completely gauge-invariant, regardless of the number of fields involved. 

The key to the argument is the correspondence between the weight-lattice for a given representation and the collection of two-cycles that simultaneously shrink at some place along the base. Recall the explicit meaning of `the gauge-invariant operator $\prod_iX_i$': it is the singlet-component of the tensor-product representation $\bigotimes_iX_i$. Now, the weights of a tensor product are nothing but the outer sum of all the weights in each component representation. And so, if $\prod_iX_i$ is completely gauge-invariant, there must be a set of weights $x_i$, one coming from each $X_i$'s weight-lattice, such that $\sum_ix_i=0$. 

In M-theory this means that as we travel from $X_1$ to $X_2$, there is a two-cycle (homology class) $[x_1]$ that vanishes at $X_1$ and grows until we reach $X_2$, at which point it joins with $[x_2]$ (which is shrunk at $X_2$) to form the class $[x_1]+[x_2]$; this in turn grows from $X_2$ to $X_3$, and so on, until it becomes $\sum_{i=1}^{n-1}[x_1]$ from $X_{n-1}$ to $X_n$. But as we have just seen, gauge-invariance implies that $\sum_{i=1}^{n-1}[x_1]=-[x_n]$, and so this two-cycle will shrink again at $X_n$. This means that there is a foliated three-cycle supporting the entire collection of conical singularities, and this will generate the corresponding operator in the superpotential. 

Therefore, the structure of three-cycles in the geometry of an ALE-fibred $G_2$-manifold is fixed precisely by gauge-invariance alone---but by gauge-invariance including with respect to the complete set of additional $U_1$-factors. As such, much of the structure of these models can be understood purely in terms of effective field theory, where the spectrum of matter fields descends from the branching of ADE-group adjoints\footnote{The branching of an adjoint---being real---always results in vector-like matter; geometrically, this corresponds to the fact that in each fibre there are two-cycles with opposite areas transforming in conjugate representations, which vanish together over the base. Therefore, there are really three-cycles for every gauge-invariant interaction among the `hypermultiplets' which descend from the branching of ADE-adjoints. But M-theory on a $G_2$-manifold, giving $\mathcal{N}=1$, will support only half of each such `hypermultiplet;' which chiral components can be massless will be discussed below.}. But because these $U_1$-factors fix the {\it geography} of conical singularities in explicit, geometrically engineered manifolds, it is possible for us to go beyond effective field theory and compute the complete superpotential explicitly (including coefficients) in terms of (the parameters of) the moduli maps $(f_1,\ldots,f_n)$. 

\subsection{Charged Chiral Matter in M-Theory and F-Theory}\label{fmdiffs2}
As we have seen, many of the na\"{i}vely non-generic structures necessary for phenomenology are in fact {\it automatically} present in both F-theory and M-theory when compactified on an ALE-fibred manifold. Nonetheless, F-theory turns out to be enormously more flexible than M-theory, making it easier to coerce into giving realistic phenomenology. The source of much of this flexibility is how chiral matter is generated in F-theory as opposed to in M-theory. 

In F-theory, which chiral components of the geometrically engineered `hypermultiplets' end up massless is determined by the $U_1$-flux supported along each matter-curve (see, e.g., \cite{Beasley:2008dc,Beasley:2008kw,Donagi:2008ca}). 
 The amount of flux along each matter-curve is tunable, and so long as we envision the base of our fibration to be sufficiently structured---such as a high-degree del Pezzo surface---there are typically enough independent (complex) one-cycles upon which we can wrap the matter-curves so that the amount of flux on each curve is {\it separately tunable}. In particular, adding no flux will project-out neither component of the hypermultiplet living along it, meaning that {\it no} massless charged matter will be supported along the curve, making exotic and unwanted matter trivial to ignore in F-theory. Additionally, one can adjust the number of units of flux to project-out multiple massless matter fields along a single curve. In general, one can obtain any number of massless chiral fields of a given representation {\it or} its conjugate along a corresponding matter-curve locally\footnote{Of course, global constraints such as anomaly cancellation and tadpole conditions must be satisfied in any compact model.}. 

The origin of massless matter in F-theory allows for a very interesting solution to the doublet-triplet splitting problem of $SU_5$-models \cite{Beasley:2008dc,Beasley:2008kw}. Ordinarily, the number of massless $\mathbf{5}$'s (or $\bar{\mathbf{5}}$'s) that live along a specific curve is set by the flux of one or more of the extra $U_1$'s coming from the typical fibre (which is almost always taken to be $\widehat{E_6}$ or higher). If one started with such a geometrically engineered model and broke $SU_5$ to the Standard Model using some canonical Higgsing mechanism, then the doublet-triplet splitting problem would remain, {\it prima facia} intact. However, in F-theory one has the option of breaking $SU_5$ to the Standard Model through internal {\it hypercharge flux}---something that would ordinarily break hypercharge in other string models, such as the heterotic string. 
 This breaking of $SU_5$ alters the `number of massless fields' flux-calculation, making the number of massless $(\mathbf{1},\mathbf{2})$'s and the number of massless $(\mathbf{3},\mathbf{1})$'s living along a $\mathbf{5}$-curve two separate calculations. As such, one can balance the competing $U_1$-fluxes so that one finds massless Higgs doublets but not their colour-triplet partners (see \cite{Beasley:2008dc,Beasley:2008kw}). 

Almost none of this freedom is available in M-theory: conical singularities always give rise to {\it one} set of massless chiral matter---there is no analogue of having matter singularities that can be ignored, or having multiplicities of matter arising from a single conical singularity; and local models in M-theory appear to have no (perturbative) analogue to F-theory's solution to the doublet-triplet splitting problem. How doublet-triplet splitting can be achieved in M-theory will be discussed in more detail when we construct concrete examples in M-theory in \mbox{Section \ref{mtheg}}.

In M-theory, the only sure way to avoid dangerous exotic matter is to ensure that their corresponding conical singularities simply do not appear in the fibration. For example, if an exotic (anti) quark-doublet $(\bar{\mathbf{3}},\mathbf{2})$ were to live at the solution to \mbox{$5Y(t)=0$}, the only clear way to exclude it from the spectrum is to make $Y(t)$ a constant map, independent of the base. This kind of `parallel-projection' can also be used in F-theory, but would seem to be unnecessary. One of the reasons why M-theory models are so constrained is that parallelization is often the only way to project-out unwanted matter from the theory, and the constraints used to do this often end up excluding models that would be otherwise good in F-theory.%

The one freedom of F-theory that seems to be allowed in M-theory is to independently choose the relative conjugations of fields living at disparate conical singularities. This contradicts the overconfident claims of the authors of \cite{Bourjaily:2008ji}, and deserves a more detailed explanation, to be found in a later note \cite{Bourjaily:2009aa}. 

\subsection{Achieving Realistic Superpotentials in F-Theory and M-Theory}\label{yukdiffs}
We have described how interactions in the superpotential arise in F-theory and M-theory through the natural geometric structures present in ALE-fibrations. Recall that the coefficients of these operators in F-theory, while not necessarily easy to compute, are not typically very hierarchical. For M-theory in contrast, all the interaction coefficients come from instantons and are therefore exponentially suppressed but are easily computed. This may seem to spell disaster for any model in M-theory, because any purely-exponential superpotential is in obvious conflict, for example, with the top-quark's Yukawa coupling. How can this be remedied?

One possible solution in M-theory is as follows. While a typical M-theory fibration will have many non-vanishing three-cycles, it is always possible to tune moduli so that any {\it one}\footnote{It may sometimes be possible to enforce more than one Yukawa coupling to be $\mathcal{O}(1)$; but such a constraint is usually either inconsistent with necessary parallelization choices, or ends up leading to enhanced gauge symmetries.} of these three-cycles shrinks to zero size, forcing a single Yukawa coupling to be $\mathcal{O}(1)$. Graphically, this corresponds to letting the M-theory `slice' pass through a triple-intersection of matter-curves in the F-theory plane (see \mbox{Figure \ref{e7_exempli}}). It is therefore possible in M-theory to generate a superpotential dominated by single large Yukawa coupling which is otherwise hierarchical. This could be the explanation of the unusually unbalanced flavour structure of the Standard Model.

But it may not be necessary to tune moduli at all to generate an $\mathcal{O}(1)$ operator in the superpotential in M-theory: there may be dynamics capable of generating such couplings. For example, large effective-operators can be generated by quartic couplings in the superpotential involving Standard Model-singlet fields which acquire large vacuum expectation values. This can arise through vacuum-realignment, and we will see how this happens in \mbox{Section \ref{mtheg}}. Another possible way to get a large Yukawa coupling, also due to vacuum realignment, is more analogous to the see-saw mechanism of neutrino physics; models exploiting this possibility will be discussed in a later note \cite{Bourjaily:2009ab}.

But an order-one Yukawa coupling in the up-type quark sector is not itself sufficient to generate a realistic spectrum of quark masses. Indeed, as pointed out in \mbox{Ref.\ \cite{Beasley:2008kw}}, there is a generic impediment to getting a realistic spectrum of quark masses in geometrically engineered models: cubic interactions in the superpotential arising from triple-intersections of matter curves or M2-brane instantons should always connect apparently {\it distinct} matter fields. So, for example, we expect an up-type Higgs field $\mathbf{5}^{H}$ to always couple to distinct matter-fields $\mathbf{10}_a$ and $\mathbf{10}_b$ according to \mbox{$W\supset \lambda\,\mathbf{10}_a\,\mathbf{10}_b\,\mathbf{5}^H$}; this would lead to a mass matrix of the form,\vspace{-0.2cm}\begin{equation}M^2_{u}=\left(\begin{array}{ccc}0&A&B\\A&0&C\\B&C&0\end{array}\right);\vspace{-0.2cm}\end{equation}
but any such matrix has the property that $\mathrm{det}\left(M^2_u\right)=\mathrm{tr}\left(M^2_u\right)=0$, and therefore cannot be dominated by a single large eigenvalue. 

In F-theory, this problem can be solved by assuming that among the three {\it apparently} distinct curves that intersect to generate a $\mathbf{10}_a\,\mathbf{10}_b\,\mathbf{5}^H$ coupling, the two $\mathbf{10}$-curves are in fact the same globally; that is, the triple-intersection in fact involves a self-intersection of one of the matter-curves. Such a situation would generate additional `on-diagonal' interactions in the superpotential, avoiding the obstruction noted above.  

One technical problem with relying on self-intersections in F-theory, however, is that it is not possible to see such global structures within a single local patch of a fibration (which is what we discuss in this paper), unless one uses higher-degree maps from the base into the moduli space. Both to avoid the arbitrariness of such maps, and to treat F-theory and M-theory in parallel, we prefer not to rely on self-intersections.


Luckily, there are at least two alternative mechanisms that can lead to self-couplings in the superpotential, which can work in both F-theory and M-theory; we will see examples of each in \mbox{Section \ref{exempliModels}}. The method that we will use in our F-theory examples of \mbox{Section \ref{ftheg}} was mentioned in Ref.\ \cite{Heckman:2008ads}, and requires the existence of matter-singularities that are multiply-enhanced in rank relative to the base. Consider for example the two-rank enhancement of $SU_5\times U_1^a\times U_1^b\to SO_{12}$, where matter living along the $SO_{12}$-curve transforms according to the non-adjoint components of\vspace{-0.2cm}\begin{equation}\begin{split}\mathbf{66}=&\mathbf{24}_{0,0}\oplus\mathbf{1}_{0,0}\oplus\mathbf{1}_{0,0}\oplus\mathbf{10}_{1,\text{-}1}\oplus\mathbf{5}_{\text{-}2,2}\oplus\bar{\mathbf{5}}_{\text{-}3,3}\\
&\hspace{90.5pt}\oplus\bar{\mathbf{10}}_{\text{-}1,1}\oplus\bar{\mathbf{5}}_{2,\text{-}2}\oplus\mathbf{5}_{3,\text{-}3}\,.\end{split}\quad\vspace{-0.2cm}\end{equation}
Here, all the matter fields would live along the same curve, $a-b=0$. 
This could generate the direct, self-interaction $\mathbf{10}_{1,\text{-}1}\,\,\mathbf{10}_{1,\text{-}1}\,\,\mathbf{5}_{\text{-}2,2}$, contributing diagonally to the superpotential. This is precisely the structure that we use in \mbox{Section \ref{ftheg}} to engineer a concrete realization of Ref.\ \cite{Heckman:2008ads}'s `Diamond Ring' model in F-theory. We anticipate that this type of mechanism should also work in M-theory---and because such a singularity would automatically generate an on-diagonal interaction---presumably with an $\mathcal{O}(1)$-coefficient---it would cure two problems at once. 

The second way we can generate on-diagonal Yukawa couplings is to rely on effective operators, generated by quartic (or higher) terms in the superpotential involving singlet fields which develop large vacuum expectation values. This is the method we use in the M-theory example discussed in \mbox{Section \ref{mtheg}}. Like above, these effective operators can sometimes be `on-diagonal' (as they are in our explicit example) and are possibly of order-one---again, solving two problems at once. The advantage of relying on higher-order operators in M-theory is that it requires no additional constraints on the (already highly-restricted) moduli space of the fibration.  

\subsection{The Necessity of $\widehat{E_8}$?}\label{neede8}
From our analysis of the branching of $\widehat{E_7}$ in \mbox{Section \ref{sectexempli}}, it is not hard to see why $\widehat{E_7}$ is inadequate for complete phenomenology in M-theory: although there are three conical singularities supporting $\mathbf{10}$-dimensional representations, there are only four supporting $\mathbf{5}$-dimensional ones. Because multiple copies of a representation cannot come from a single conical singularity in M-theory, there are not enough $\mathbf{5}$-dimensional representations to account for three generations of matter together with two Higgs fields. Breaking to the gauge group $SU_3\times SU_2$ will not help either, because the adjoint of $SU_5$ will supply only an additional bi-fundamental field, and the model would still lack the requisite additional $SU_2$-doublet. 

In F-theory, however, the issue is much more subtle. To begin with, we can imagine having any of the matter-curves in Table \ref{exempli_table} support multiple generations, and so the problem above is easily avoided. Rather, the problem of an $\widehat{E_7}$-fibration in F-theory has more to do with the variety of couplings simultaneously possible in the superpotential. The problem that we will find with this setup is that no choice of fluxes will allow for all the types of operators needed for phenomenology. Let us see how this happens.

Of the four $\mathbf{5}$-dimensional representations listed in \mbox{Table \ref{exempli_table}}, only $F_4$ is unable to play the role of the up-type Higgs because it has no possible coupling with two $T_i$'s. But any one of $\bar{F}_{1,2,3}$ can play this role; because each is on equal footing, we may without loss of generality choose to take $\bar{F_1}$ to be the up-type Higgs $\mathbf{5}$, with an interaction of the form $T_1\,\bar{F_1}\,T_2$. Let us now identify the down-type Higgs field. It must couple to matter according to $\bar{\mathbf{5}}{\!\!\!\phantom{\mathbf{5}}}^H\bar{\mathbf{5}}{\!\!\!\phantom{\mathbf{5}}}^M\mathbf{10}$, and must also appear in a coupling of the form ${\mathbf{5}}^H\bar{\mathbf{5}}{\!\!\!\phantom{\mathbf{5}}}^H\mathbf{1}^{(\dag)}$ in order to dynamically generate a $\mu$-term (either through the superpotential or the K\"{a}hler potential). The possibility of a dynamical $\mu$-term requires that $\bar{\mathbf{5}}{\!\!\!\phantom{\mathbf{5}}}^H$ be either $F_2$ or $F_3$; in either case, the only possibility for the matter-field $\bar{\mathbf{5}}{\!\!\!\phantom{\mathbf{5}}}^M$ is $\bar{F_4}$. But then, notice that there is no $SU_5$-singlet available to play the role of a right-handed neutrino in a coupling of the form \mbox{$\bar{\mathbf{5}}{\!\!\!\phantom{\mathbf{5}}}^H\bar{\mathbf{5}}{\!\!\!\phantom{\mathbf{5}}}^M\mathbf{1}^{\nu^c}$}. This may not be fatal, but it is unlikely to be realistic. And so, even F-theory on general $\widehat{E_7}$-fibred manifold with arbitrary fluxes is not flexible enough to generate a realistic model.

But we have not yet ruled out the possibility of a non-generic $\widehat{E_7}$-fibration. Recall from our discussions regarding self-interactions that one can {\it actively engineer} additional couplings by forcing a matter-curve to be multiply-enhanced in rank. What we mean is the following. From \mbox{Table \ref{exempli_table}}, we notice that there is no way to have the field $F_4$ couple to any two of the $\mathbf{10}$'s. One operator of this form would be $T_1\,T_1\,F_4$, which is not gauge-invariant because it transforms under \mbox{$U_1^a\times U_1^b\times U_1^c$} with charges $(3,1,1)$. We could, however, guarantee such a coupling\footnote{One should be cautious about making such restrictions: it is not uncommon to find that these conditions inadvertently increase the rank of the gauge group.} by simply forcing the \mbox{{\it geometric}} condition \mbox{$c=-3a-b$}, thereby making $T_1\,T_1\,F_4$ manifestly gauge-invariant. 

Upon doing this, the resulting matter is given in the left-hand side of \mbox{Table \ref{e7egpinch}}. In this geometry, the curve $a=0$ supports a singularity of type $SO_{12}$, along which live the fields $T_1,\,F_3$ and $F_4$. And in striking contrast to Figure \ref{e7_exempli}, the geometry has {\it only a single, multifold-intersection}. 

\newcommand{\crowFF}[5]{$#1$&$\mathbf{#2}$&\ifnum#3<0$\!\!$#3\else#3\fi&\ifnum#4<0$\!\!$#4\else#4\fi&$#5=0$\\}

But of the complete set of matter-curves on the left-hand side of \mbox{Table \ref{e7egpinch}}, only six of them are necessary to generate a somewhat realistic set of couplings. By choosing appropriate fluxes, one can realize a model with the matter listed on the right-hand side of \mbox{Table \ref{e7egpinch}}, which would have a superpotential of the form, \vspace{-0.2cm}
\begin{equation}W=\lambda_1^{ij} T_i\,T_j\,H^u\,\,+\,\,\lambda_2^{ij} T_i\,M_j\,H^d\,\,+\,\,\lambda_3^{ij}M_i\nu_j^c\,H^u\,\,+\,\,\lambda_4H^u\,H^d\,S_{\mu}.\label{e7real}\vspace{-0.2cm}\end{equation}
This is perhaps the `simplest' model with three generations we can find in F-theory. Nonetheless, we do not expect that it is sufficiently flexible to reflect reality. 

\begin{table*}[!h]\begin{center}
\begin{tabular}{|rlll|r|}
\hline
&$\!\!SU_5\,\times\,\!\!\!\!$&$\!U_1^a\,\times\!\!\!\!$&$\!U_1^b$&Location\\\hline
\crowFF{T_1}{10}{1}{0}{a}
\crowFF{T_2}{10}{0}{1}{b}
\crowFF{T_3}{10}{-3}{-1}{-3a-b}
\crowFF{F_1}{\bar{5}}{1}{1}{a+b}
\crowFF{F_2}{\bar{5}}{-2}{-1}{-2a-b}
\crowFF{F_3}{\bar{5}}{-3}{0}{-3a}
\crowFF{F_4}{5}{-2}{0}{-2a}
\crowFF{S_1}{1}{1}{-1}{a-b}
\crowFF{S_2}{1}{4}{1}{4a+b}
\crowFF{S_3}{1}{3}{2}{3a+2b}
\hline
\end{tabular}\hspace{0.5cm}\begin{tabular}{|rlll|r|}
\hline
&$\!\!SU_5\,\times\,\!\!\!\!$&$\!U_1^a\,\times\!\!\!\!$&$\!U_1^b$&Location\\\hline
\crowFF{3\times\,\,T_i\,}{10}{1}{0}{a}
\crowFF{3\times M_i\,}{\bar{5}}{-2}{-1}{-2a-b}
\crowFF{\!H^u}{5}{-2}{0}{-2a}
\crowFF{\!H^d}{\bar{5}}{1}{1}{a+b}
\crowFF{3\times\,\,\nu_i^c}{1}{4}{1}{4a+b}
\crowFF{S_{\mu}}{1}{3}{2}{3a+2b}
\hline
\end{tabular}\caption{\label{E7_to_SU5_modified} The set of matter-curves in the fibration $\widehat{E_7}(a,b,-3a-b,0,0,0,0)$ on the left, and for a special choice of fluxes on the right---where fields have been suggestively renamed.\label{e7egpinch}}\vspace{-1cm}\end{center}\end{table*}
\newpage

\section[Exempli Gratiae: Realistic Models in F-theory and M-Theory]{Exempli Gratia: Realistic Models in F-theory and M-Theory}\label{exempliModels}
In this section, we pedagogically construct a series of examples which constitute an existence proof of purely local, truly phenomenological models with three generations in F-theory and M-theory. Because of its greater flexibility, we will start by building models in F-theory in \mbox{Section \ref{ftheg}}, and specialize to the more constrained framework of M-theory in \mbox{Section \ref{mtheg}}. 

\subsection{Constructing Phenomenological Models in F-Theory}\label{ftheg}
Our goal of this subsection will be to construct complete phenomenological models in F-theory. We will start with the example mentioned in the introduction, an $E_6$ grand-unified model based on a general fibration of $\widehat{E_8}$, and see how this fibration can be altered, or `unfolded' in the language of \cite{Bourjaily:2007vx}, to incorporate sequentially less gauge symmetry. In particular, we will first show how to unfold the fibration giving $E_6$ to generate an $SO_{10}$-model with a Peccei-Quinn symmetry resulting from the decomposition \mbox{$E_6\supset SO_{10}\times U_1^{PQ}$}. We then further unfold this fibration to one giving mere $SU_5$ gauge symmetry, bringing us quite close to realistic phenomenology. 

The authors of Ref.\ \cite{Heckman:2008ads} qualitatively described the particle content and interactions of a particularly compelling $SU_5$ grand-unified model in F-theory---similar to those described in \cite{Marsano:2008jq}---which they called the `Diamond Ring.' This model has the potential for local supersymmetry  breaking with gauge-mediation---close in spirit to the `sweet spot' supersymmetry breaking described in \cite{Ibe:2007km}---with a rather nice resolution of the $\mu/B\mu$-problem among other phenomenologically appealing features (see \cite{Heckman:2008ads} for more details). 
While it was clear to the authors of \cite{Heckman:2008ads} that such a model was in principle possible locally in F-theory, they did not present any concrete realization of a geometry that would generate it. We will meet this challenge in \mbox{Section \ref{diamondring}}, giving a natural origin of their `Diamond Ring,' constructed as a constrained $\widehat{E_8}$-fibration. In \mbox{Section \ref{solitaire}} we obtain a less elaborate version of the `Diamond Ring,' closer in spirit to that in \cite{Heckman:2008ads}.

\subsubsection{The Local Origin of Three Generations: $E_8\to E_6\times U_1^a\times U_1^b$}\label{originofthree}
\newcommand{\TwoRes}[5]{$#1$&$\mathbf{#2}$&\ifnum#3<0$\!\!$#3\else#3\fi&\ifnum#4<0$\!\!$#4\else#4\fi&$#5=0$\\}
As described in the introduction, one potential explanation for the origin of three generations could be that any generic fibration of $\widehat{E_8}$ will include (at least) three $\mathbf{27}$'s of $E_6$-worth of matter. As we have seen in \mbox{Section \ref{sectexempli}}, this is nothing more than a reflection of the representation-theoretic fact that the adjoint of $E_8$ branches into its subgroup $E_6\times U_1^a\times U_1^b$ according to\begin{equation}\begin{split}\mathbf{248}=&\mathbf{78}_{0,0}\oplus\mathbf{1}_{0,0}\oplus\mathbf{1}_{0,0}\oplus\mathbf{27}_{1,1}\oplus\mathbf{27}_{\text{-}2,0}\oplus\mathbf{27}_{1,\text{-}1}\hspace{2.85pt}\oplus\mathbf{1}_{3,\text{-}1}\oplus\mathbf{1}_{3,1}\hspace{5.7pt}\oplus\mathbf{1}_{0,2}
\\&\hspace{90.5pt}\oplus\bar{\mathbf{27}}_{2,0}\oplus\bar{\mathbf{27}}_{\text{-}1,1}\oplus\bar{\mathbf{27}}_{\text{-}1,\text{-}1}\oplus\mathbf{1}_{\text{-}3,1}\oplus\mathbf{1}_{\text{-}3,\text{-}1}\oplus\mathbf{1}_{0,\text{-}2}.\end{split}\,\,\label{e8toe6branch}\end{equation}

\begin{figure}[!tr]
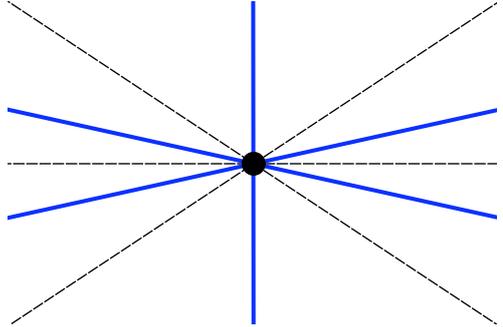
\caption{Matter curves along the $E_6$-base.\hspace{-9.5cm}$~$\label{fthe6model}}\end{figure}

Geometrically, the branching in equation (\ref{e8toe6branch}) is reflected in the fibration given by $\widehat{E_8}(a+b,a-b,0,0,0,0,0,0)$. Following a similar---yet easier---exercise to that of \mbox{Section \ref{sectexempli}}, we see that this fibration would result in the six matter-curves listed in Table \ref{e8toe6xu1u1}. In \mbox{Figure \ref{fthe6model}}, we show a realization of this geometry, where matter-curves generating $\mathbf{27}$'s of $E_6$ are drawn in solid blue and singlets are in thinly-dashed black.\vspace{-0.5cm}

\begin{table}[tl]\vspace{-1.16cm}\caption{The general matter content resulting\hspace{6.95cm}\mbox{${~}$} \hspace{10cm} from $\widehat{E_8}(a+b,a-b,0,0,0,0,0,0)$ in F-theory.\label{e8toe6xu1u1}}\hspace{0.95cm}\begin{tabular}{|rlll|r|}
\hline
$\,\qquad\!\!$&$\!\!\!E_{6}\times\,\!\!\!\!$&$\!U_1^a\times\!\!\!\!\!$&$\!U_1^b$&Location\\\hline
\TwoRes{T_1}{27}{1}{1}{a+b}
\TwoRes{T_2}{27}{1}{-1}{a-b}
\TwoRes{T_3}{27}{-2}{0}{-2a}
\TwoRes{S_1}{1}{3}{-1}{3a-b}
\TwoRes{S_3}{1}{3}{1}{3a+b}
\TwoRes{S_2}{1}{0}{2}{2b} 
\hline
\end{tabular}\hspace{1.95cm}\raisebox{-2cm}{\includegraphics[scale=1.05,angle=0]{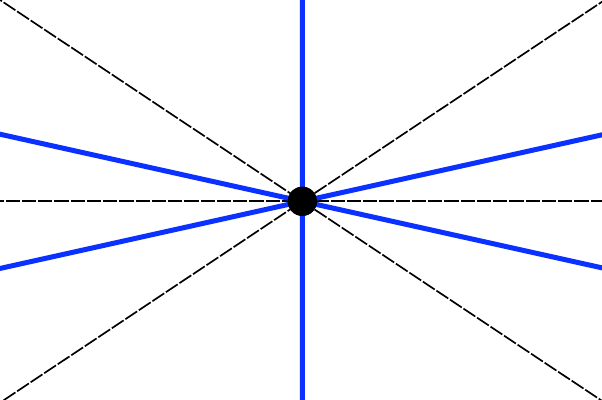}}\end{table}

Although the appearance of three interacting $\mathbf{27}$'s of $E_6$ in the the resolution $\widehat{E_8}(a+b,a-b,0,0,0,0,0,0)$ may appear to have been inserted by hand, it is actually extremely general. Any F-theory model with $E_6$ gauge symmetry and more than a single $\mathbf{27}$ of matter should possess this structure. To see why this is so, recall that each matter-curve giving rise to a $\mathbf{27}$ (solid blue lines in \mbox{Figure \ref{fthe6model}}) is a Riemann surface within a compact complex two-cycle supporting the $E_6$ gauge theory singularity. Whenever there are multiple $\mathbf{27}$-curves, they will generically intersect---being curves in a compact two-fold. And, because the singularity at the intersection of two $\mathbf{27}$-curves must be of rank-8, be simple, and contain $\widehat{E_6}$, it must be $\widehat{E_8}$ (see \mbox{Appendix \ref{TIeqGIO}}). Therefore, we see again both the ubiquity of $\widehat{E_8}$-fibrations, and how $E_8$ representation theory automatically leads us to consider a theory with (at least) {\it three} generations of matter. 

\subsubsection{Unfolding $E_6$ to Generate an $SO_{10}$ Grand-Unified Model}\label{FTheorySO10model}
By analogy with \mbox{Section \ref{sectexempli}}, it should be easy to guess how the $\widehat{E_8}$-fibration above can be `unfolded' to one with less than $E_6$-gauge symmetry. For example, the fibration \mbox{$\widehat{E_8}(a+b,a-b,c,0,0,0,0,0)$}, where $a,b,$ and $c$ are arbitrary maps of the base, will automatically preserve a singularity of type $E_5(\equiv SO_{10})$ everywhere over the base, generating $SO_{10}$ gauge theory. Of course, any deformation of $E_6$ to $SO_{10}$ should suffice, but the explicit forms of the moduli functions $(f_1,\ldots,f_8)$ in terms of the maps $a,b,$ and $c$ reflects a choice of basis for the $U_1$-factors in the decomposition \mbox{$E_8\supset SO_{10}\times U_1\times U_1\times U_1$}, and some choices are more convenient than others. 

Perhaps the most phenomenologically well-motivated choice is the familiar decomposition $E_6\supset SO_{10}\times U_1^{PQ}$, where $U_1^{PQ}$ is the Peccei-Quinn symmetry, useful both for solving the strong-CP problem and to prevent proton decay in supersymmetric grand-unified models\footnote{Recall that matter-parity can be embedded as a discrete subgroup of $U_1^{PQ}$, \cite{Farrar:1978R}.} \cite{Peccei:1977ur,Peccei:1977hh}. After our practice in \mbox{Section \ref{sectexempli}}, it is not too hard to find an explicit presentation of the fibration which highlights the $U_1^{PQ}$ subgroup of $E_6$; one such example is the fibration\vspace{-0.0cm}\begin{equation}\widehat{E_8}(a+b+c,a-b+c,-c,-c,-c,-c,-c,2c),\label{e8toso10}\vspace{-0.0cm}\end{equation}
where $U_1^{PQ}\equiv U_1^c$. To see that this fibration preserves an $SO_{10}$-singularity regardless of the maps $a,b,$ and $c$, one should check that the set of two-cycles labelled by\begin{equation}\left\{\begin{array}{c}\hspace{1.25cm}(-e_0+e_6+e_7+e_8)\\(e_3-e_4)\,\,(e_4-e_5)\,\,(e_5-e_6)\,\,(e_6-e_7)\end{array}\right\}\label{so10roots}\end{equation} all vanish, generate the $SO_{10}$ root-lattice, and are arranged in $\widehat{E_8}$ according to the $SO_{10}$ Dynkin diagram (as they are in (\ref{so10roots})). 

The set of matter-curves present in this fibration is given in \mbox{Table \ref{fthso10model}}. Notice that while we started with only three $\mathbf{27}$'s of matter in $E_6$, we find in the $SO_{10}$ model an additional, `exotic' $\bar{\mathbf{16}}$ (or $\mathbf{16}$) of matter living along the curve given by $\mp3c=0$. This descends from the adjoint decomposition of $E_6\to SO_{10}\times U_1^c$: \begin{equation}\mathbf{78}=\mathbf{45}_0\oplus\mathbf{1}_0\oplus\mathbf{16}_{3}\oplus\bar{\mathbf{16}}_{-3}.\end{equation}

This exotic matter could be excluded simply by insisting that \mbox{$c(t)$} were a non-vanishing constant, independent of the base; notice that such a condition would not exclude any of the other matter in \mbox{Table \ref{fthso10model}}. This kind of `parallel-projection' is critical in M-theory where it is often the only tool available to exclude exotics, but it is a bit superfluous in F-theory where matter-curves without flux do not generate massless chiral matter. 
\newcommand{\FtheoryMOT}[6]{$#1$&$\mathbf{#2}$&\ifnum#3<0$\!\!$#3\else#3\fi&\ifnum#4<0$\!\!$#4\else#4\fi&\ifnum#5<0$\!\!$#5\else#5\fi&$#6=0$\\}
\begin{table}[b]\begin{center}
\vspace{-0.4cm}\begin{tabular}{|rllll|r|}
\hline
$\,\qquad\!\!$&$\!\!\!SO_{10}\times\,\!\!\!\!$&$\!U_1^a\times\!\!\!\!\!$&$\!U_1^b\times\!\!\!\!\!$&$\!U_1^c$&Location\\\hline
\FtheoryMOT{T_1}{16}{1}{1}{-1}{a + b - c}
\FtheoryMOT{T_2}{16}{1}{-1}{-1}{a - b - c}
\FtheoryMOT{T_3}{16}{-2}{0}{-1}{-2 a - c}
\FtheoryMOT{T_X^c\!\!}{\bar{16}}{0}{0}{-3}{-3 c}
\FtheoryMOT{H}{10}{1}{1}{2}{a + b + 2 c}
\FtheoryMOT{Y_a}{10}{1}{-1}{2}{a - b + 2 c}
\FtheoryMOT{Y_b}{10}{-2}{0}{2}{-2 a + 2 c}
\FtheoryMOT{X_1}{1}{-1}{-1}{4}{-a - b + 4 c}
\FtheoryMOT{X_2}{1}{-1}{1}{4}{-a + b + 4 c}
\FtheoryMOT{N_2^c}{1}{-2}{0}{-4}{-2 a - 4 c}
\FtheoryMOT{N_3^c}{1}{3}{-1}{0}{3 a - b}
\FtheoryMOT{S_1}{1}{3}{1}{0}{3 a + b}
\FtheoryMOT{S_2}{1}{0}{2}{0}{2 b}
\hline
\end{tabular}\caption{The matter-curves generated by the fibration specified by (\ref{e8toso10}), named with foresight toward later subsections. Here, $U_1^c\equiv U_1^{PQ}$.\label{fthso10model}}\end{center}\vspace{-1.5cm}\end{table}
\newpage

\begin{figure}[h]\begin{center}\includegraphics[scale=1.755,angle=0]{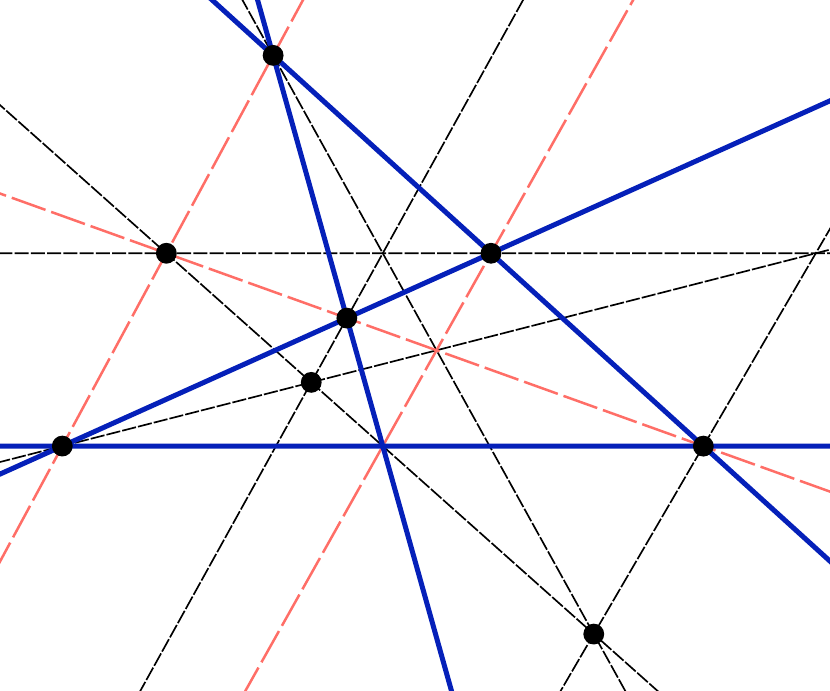}\caption{An illustration of the fibration (\ref{e8toso10}) which generates a canonical $SO_{10}$-model in F-theory with matter indicated in Table \ref{fthso10model}. Here, thick, solid blue lines represent $\mathbf{16}$'s and $\bar{\mathbf{16}}$'s, widely-dashed red lines are $\mathbf{10}$'s, and the finely-dashed black lines are $SO_{10}$-singlets. Black dots indicate triple-intersections that generate gauge-invariant Yukawa couplings based on the choice of fluxes generating the conjugations indicated in \mbox{Table \ref{fthso10model}}.\label{fthso10modelslice}}\end{center}\end{figure}

An example of the fibration (\ref{e8toso10}) is illustrated in \mbox{Figure \ref{fthso10modelslice}}, where we have taken the maps $a,b,$ and $c$ to be random, linear functions of a two-parameter base. In that figure, the thick, solid blue lines represent $\mathbf{16}$-dimensional representations of $SO_{10}$, widely-dashed red lines are $\mathbf{10}$'s, and thinly-dashed black lines indicate $SO_{10}$-singlets. Adding $\pm1$ unit of flux along each matter-curve so as to generate the conjugations listed in \mbox{Table \ref{fthso10model}} will determine which triple-intersections generate gauge-invariant operators among {\it massless} chiral fields in the superpotential; only these triple-intersections have been indicated by dots in \mbox{Figure \ref{fthso10modelslice}}. \\~\\~\\~\\

\newpage

\subsubsection{A Diamond in the Rough: Unfolding $SU_5$ Out of $SO_{10}$}\label{fthsu5canon}
\newcommand{\FtheoryZT}[7]{$#1$&$\mathbf{#2}$&\ifnum#3<0$\!\!$#3\else#3\fi&\ifnum#4<0$\!\!$#4\else#4\fi&\ifnum#5<0$\!\!$#5\else#5\fi&\ifnum#6<0$\!\!$#6\else#6\fi&$#7=0$\\}
Let us now further reduce the degree of gauge symmetry generated by our fibration by allowing an even more general set of maps from the base into the $\widehat{E_8}$ moduli. Looking at the basis of shrunk two-cycles given in (\ref{so10roots}), we see that blowing up either $(-e_0+e_6+e_7+e_8)$ or $(e_6-e_7)$ will leave us with an $SU_5$ sub-lattice of shrunk two-cycles. Motivated by the familiar embedding of $SU_5\times U_1^d\subset SO_{10}$, we are led to define our fibration by \vspace{-0.2cm}\begin{equation}\widehat{E_8}(a+b+c+d,a-b+c+d,-c-d,-c-d,-c-d,-c-d,-c+3d,2c-2d);\label{e8tosu5resolution}\vspace{-0.2cm}\end{equation}
notice that this preserves the vanishing of the two-cycle labelled by $(-e_0+e_6+e_7+e_8)$, but blows-up that of $(e_6-e_7)$.

Following the same analysis as before, we can directly determine the complete set of matter-curves present in the fibration specified by (\ref{e8tosu5resolution}). This is listed in \mbox{Table \ref{fthsu5modelall}}, where we have divided the matter-curves into those along which we will and will not project-out chiral matter, written on the left- and right-hand sides, respectively.\nopagebreak
\begin{table}[!h]
\vspace{-0.2cm}\hspace{-1.3cm}\begin{tabular}{|rlllll|r|}
\hline
$\,\qquad\!\!$&$\!\!\!SU_{5}\times\,\!\!\!\!$&$\!U_1^a\times\!\!\!\!\!$&$\!U_1^b\times\!\!\!\!\!$&$\!U_1^c\times\!\!\!\!\!$&$\!U_1^d$&Location\\\hline
\FtheoryZT{T_1}{10}{1}{1}{-1}{-1}{a + b - c - d}
\FtheoryZT{T_2}{10}{1}{-1}{-1}{-1}{a - b - c - d}
\FtheoryZT{T_3}{10}{-2}{0}{-1}{-1}{-2 a - c - d}
\FtheoryZT{M_1}{\bar{5}}{1}{1}{-1}{3}{a + b - c + 3 d}
\FtheoryZT{M_2}{\bar{5}}{1}{-1}{-1}{3}{a - b - c + 3 d}
\FtheoryZT{M_3}{\bar{5}}{-2}{0}{-1}{3}{-2 a - c + 3 d}
\FtheoryZT{H^u}{5}{1}{1}{2}{2}{a + b + 2 c + 2 d}
\FtheoryZT{H^d}{\bar{5}}{1}{1}{2}{-2}{a + b + 2 c - 2 d}
\FtheoryZT{Y_1}{5}{-1}{1}{-2}{2}{-a + b - 2 c + 2 d}
\FtheoryZT{Y_1^c\!}{\bar{5}}{2}{0}{-2}{-2}{2 a - 2 c - 2 d}
\FtheoryZT{Y_2}{5}{2}{0}{-2}{2}{2 a - 2 c + 2 d}
\FtheoryZT{Y_2^c\!}{\bar{5}}{-1}{1}{-2}{-2}{-a + b - 2 c - 2 d}
\FtheoryZT{X_1}{1}{-1}{-1}{4}{0}{-a - b + 4 c}
\FtheoryZT{X_2}{1}{-1}{1}{4}{0}{-a + b + 4 c}
\FtheoryZT{\nu_1^c}{1}{1}{1}{-1}{-5}{a + b - c - 5 d}
\FtheoryZT{\nu_2^c}{1}{1}{-1}{-1}{-5}{a - b - c - 5 d}
\FtheoryZT{\nu_3^c}{1}{-2}{0}{-1}{-5}{-2 a - c - 5 d}
\FtheoryZT{N_1^c}{1}{0}{0}{-3}{5}{-3 c + 5 d}
\FtheoryZT{N_2^c}{1}{-2}{0}{-4}{0}{-2 a - 4 c}
\FtheoryZT{N_3^c}{1}{3}{-1}{0}{0}{3 a - b}
\hline
\end{tabular}\begin{tabular}{|rlllll|r|}
\hline
$\,\qquad\!\!$&$\!\!\!SU_{5}\times\,\!\!\!\!$&$\!U_1^a\times\!\!\!\!\!$&$\!U_1^b\times\!\!\!\!\!$&$\!U_1^c\times\!\!\!\!\!$&$\!U_1^d$&Location\\\hline
\FtheoryZT{T_X^c\!\!}{\bar{10}}{0}{0}{-3}{1}{-3 c + d}
\FtheoryZT{T_X\!\!}{10}{0}{0}{0}{4}{4 d}
\FtheoryZT{M_X^c\!\!}{5}{0}{0}{-3}{-3}{-3 c - 3 d}
\FtheoryZT{S_1}{1}{3}{1}{0}{0}{3 a + b}
\FtheoryZT{S_2}{1}{0}{2}{0}{0}{2 b}
\hline
\end{tabular}\caption{The complete set of matter-curves generated by the fibration (\ref{e8tosu5resolution}), separated according to whether or not we desire to add flux along them. On the left are the curves along which we will add flux to generate the indicated massless matter, while on the right are the curves along which we will not add flux.\label{fthsu5modelall}}\end{table}
\newpage
Recall from above that a general $\widehat{E_8}$-resolution giving $SO_{10}$ gauge symmetry will support {\it four} $\mathbf{16}$'s of $SO_{10}$---three coming from the $\mathbf{27}$'s of $E_6$, and one from the $E_6$-adjoint. Similarly, when unfolding $SO_{10}$ to $SU_{5}$, the adjoint of $SO_{10}$  can contribute an additional $\mathbf{10}$ (or $\bar{\mathbf{10}}$) of $SU_5$ to the spectrum, labelled $T_X$ in \mbox{Table \ref{fthsu5modelall}}. This additional matter is unlikely to be useful for phenomenology (but see \cite{Bourjaily:2009ab}), and so we prefer that these curves do not support any massless chiral matter. The other curves listed on the right-hand side of \mbox{Table \ref{fthsu5modelall}}, $S_{1,2}$, are less troublesome, but we will nevertheless choose them to be absent from the massless spectrum.

\begin{figure}[!b]\vspace{-2cm}\begin{center}\includegraphics[width=15cm,height=11.5cm,angle=0]{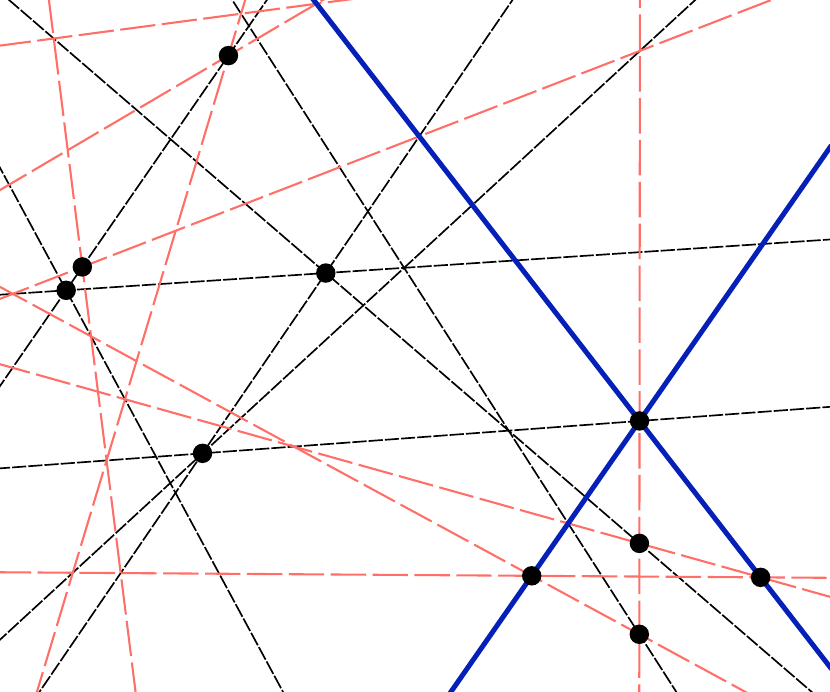}\caption{The set of matter-curves listed on the left-hand side of \mbox{Table \ref{fthsu5modelall}} resulting from the fibration (\ref{e8tosu5resolution}). $\mathbf{10}$-dimensional representations are coloured in solid blue, $\mathbf{5}$'s and $\bar{\mathbf{5}}$'s in widely-dashed red, and $SU_5$-singlets are in finely-dashed black.\label{fthsu5model1slice}}\vspace{-2cm}\end{center}\end{figure}

The set of matter-curves generating massless charged matter is illustrated in \mbox{Figure \ref{fthsu5model1slice}} for a randomly chosen set of maps $a,b,c,$ and $d$. Each triple-intersection which generates a term in the superpotential is indicated with a black dot. Suppressing coefficients, the form of the superpotential generated by this model is,\vspace{-0.2cm}
\begin{align}
\hspace{-0.75cm}W=&\,\,\,\,\,T_2\,T_3\,H^u\,\,+\,\,T_2\,M_3\,H^d\,\,+\,\,T_3\,M_2\,H^d\,\,+\,\,H^u\,M_2\,\nu _3^c\,\,+\,\,H^u\,M_3\,\nu _2^c\,\,+\,\,X_1\,Y_1\,Y_1^c\nonumber\\&\!\!+X_1\,Y_2\,Y_2^c\,\,+\,\,X_1\,N_1^c\,\nu _1^c\,\,+\,\,X_2\,N_1^c\,\nu _2^c\,\,+\,\,X_2\,N_2^c\,N_3^c.\label{fthw0}
\end{align}

This superpotential features every type of coupling necessary for phenomenology, including those that could potentially communicate supersymmetry breaking: if $X_1$ were to acquire an F-term, the fields $Y_1\oplus Y_1^c$ and $Y_2\oplus Y_2^c$ will act as messengers of this to the Standard Model through gauge interactions. 

\newpage
Nevertheless, closer inspection shows that the interactions listed in (\ref{fthw0}) are themselves insufficient for realistic phenomenology. One of the principle problems with the model we have constructed so far is that the field $T_1$ does not appear in the superpotential at all, making the up-type quark Yukawa couplings far from viable. It turns out that this problem, along with several others, can be solved at once by imposing a single constraint on the moduli functions $a,b,c,$ and $d$, similar to that introduced at the end of our discussion in Section \ref{neede8}.

\vspace{-0.2cm}\subsubsection{Cutting into Moduli Space: From Rough Diamond to Gemstone}\label{diamondring}
\newcommand{\FtheoryTT}[6]{$#1$&$\mathbf{#2}$&\ifnum#3<0$\!\!$#3\else#3\fi&\ifnum#4<0$\!\!$#4\else#4\fi&\ifnum#5<0$\!\!$#5\else#5\fi&$#6=0$\\}
Starting with the model above, let us find a condition on the moduli to enforce the gauge-invariance of the operator $T_1\,T_1\,H^u$. Notice that achieving this would not only solve our problem about generating an up-like Yukawa coupling for the field $T_1$, but it will also supply the phenomenologically important {\it on-diagonal} coupling in the up-quark mass matrix (see Section \ref{yukdiffs}). From \mbox{Table \ref{e8tosu5resolution}}, we see that the operator $T_1T_1H^u$ transforms under \mbox{$U_1^a\times U_1^b\times U_1^c\times U_1^d$} with charges $(3,3,0,0)$. That is, the singularities differ by the map $3a+3b$, which is generically non-vanishing. However, if we impose the {\it geometric} condition $b=-a$, the operator $T_1\,T_1\,H^u$ will automatically be `gauge-invariant.' This condition, when applied to (\ref{e8tosu5resolution}), leads to the fibration, \vspace{-0.3cm}\begin{equation}\widehat{E_8}(c+d,2a+c+d,-c-d,-c-d,-c-d,-c-d,-c+3d,2c-2d).\label{e8tosu5pinch}\vspace{-0.3cm}\end{equation}
\vspace{0.0cm}

\begin{table}[h]
\vspace{-0.8cm}\hspace{-0.6cm}\begin{tabular}{|rllll|r|}
\hline
$\,\qquad\!\!$&$\!\!\!SU_{5}\times\,\!\!\!\!$&$\!U_1^a\times\!\!\!\!\!$&$\!U_1^c\times\!\!\!\!\!$&$\!U_1^d$&Location\\\hline
\FtheoryTT{T_1}{10}{0}{-1}{-1}{-c - d}
\FtheoryTT{T_2}{10}{2}{-1}{-1}{2 a - c - d}
\FtheoryTT{T_3}{10}{-2}{-1}{-1}{-2 a - c - d}
\FtheoryTT{M_1}{\bar{5}}{0}{-1}{3}{-c + 3 d}
\FtheoryTT{M_2}{\bar{5}}{2}{-1}{3}{2 a - c + 3 d}
\FtheoryTT{M_3}{\bar{5}}{-2}{-1}{3}{-2 a - c + 3 d}
\FtheoryTT{H^u}{5}{0}{2}{2}{2 c + 2 d}
\FtheoryTT{H^d}{\bar{5}}{0}{2}{-2}{2 c - 2 d}
\FtheoryTT{Y_1}{5}{-2}{-2}{2}{-2 a - 2 c + 2 d}
\FtheoryTT{Y_1^c\!}{\bar{5}}{2}{-2}{-2}{2 a - 2 c - 2 d}
\FtheoryTT{Y_2}{5}{2}{-2}{2}{2 a - 2 c + 2 d}
\FtheoryTT{Y_2^c\!}{\bar{5}}{-2}{-2}{-2}{-2 a - 2 c - 2 d}
\FtheoryTT{X_1}{1}{0}{4}{0}{4 c}
\FtheoryTT{X_2}{1}{-2}{4}{0}{-2 a + 4 c}
\FtheoryTT{\nu_1^c}{1}{0}{-1}{-5}{-c - 5 d}
\FtheoryTT{\nu_2^c}{1}{2}{-1}{-5}{2 a - c - 5 d}
\FtheoryTT{\nu_3^c}{1}{-2}{-1}{-5}{-2 a - c - 5 d}
\FtheoryTT{N_1^c}{1}{0}{-3}{5}{-3 c + 5 d}
\FtheoryTT{N_2^c}{1}{-2}{-4}{0}{-2 a - 4 c}
\FtheoryTT{N_3^c}{1}{4}{0}{0}{4 a}
 \hline
\end{tabular}\begin{tabular}{|rllll|r|}
\hline
$\,\qquad\!\!$&$\!\!\!SU_{5}\times\,\!\!\!\!$&$\!U_1^a\times\!\!\!\!\!$&$\!U_1^c\times\!\!\!\!\!$&$\!U_1^d$&Location\\\hline
\FtheoryTT{T_X^c\!\!}{\bar{10}}{0}{-3}{1}{-3 c + d}
\FtheoryTT{T_X\!\!}{10}{0}{0}{4}{4 d}
\FtheoryTT{M_X^c\!\!}{5}{0}{-3}{-3}{-3 c - 3 d}
\FtheoryTT{S_1}{1}{2}{0}{0}{2 a}
\hline
\end{tabular}\caption{The matter content of the resolution (\ref{e8tosu5pinch}), corresponding to that of \mbox{Table \ref{fthsu5modelall}} upon imposing the condition $b=-a$.\label{fthsu5model2}}\vspace{-3cm}\end{table}
\newpage 
\noindent This fibration would generate the matter listed in \mbox{Table \ref{fthsu5model2}}, which is just that of \mbox{Table \ref{fthsu5modelall}} upon imposing $b=-a$. As before, we will choose fluxes so that only the fields of the left-hand side of \mbox{Table \ref{fthsu5model2}} generate massless chiral matter. Notice that the matter content is strikingly similar to that described as the `Diamond Ring' model in Ref.\ \cite{Heckman:2008ads}. 

It should be noted that the matter-curves generating $T_1$ and $H^u$ (and $M_X$) are actually the same: from \mbox{Table \ref{fthsu5model2}}, $T_1$ lives along the curve $-c-d=0$ while $H^u$ lives along $2c+2d=0$. In addition to these, the field $M_X^c$ from the right-hand side of \mbox{Table \ref{fthsu5model2}} also lives along this curve, specifically along \mbox{$-3c-3d=0$}. It is not obvious that we can choose $U_1$-fluxes along this matter-curve so as to be able to project-out both $T_1$ and $H^u$ while not also projecting out $M_X^c$. Like in Ref.\ \cite{Heckman:2008ads}, we suspect this is possible, but will not attempt to demonstrate this here. 

Along this particular curve, $-c-d=0$, we have have an especially strong enhancement of the $SU_5$-singularity typical over the base. Indeed, setting $c=-d$ in (\ref{e8tosu5pinch}) it is straight-forward to see that there is an entire $SO_{12}$-lattice of shrunk cycles along the curve. This was one of the ways discussed in \mbox{Section \ref{yukdiffs}} to generate an on-diagonal coupling: the two-rank enhancement \mbox{$SU_5\times U_1\times U_1\to SO_{12}$} allows for a direct coupling between the $\mathbf{10}$ and $\mathbf{5}$ which live along it. Interestingly, a two-rank enhanced conical singularity in M-theory we expect would result in a similar coupling, where the coefficient would also be $\mathcal{O}(1)$.

Although we imposed a restriction on moduli in order to generate the coupling $T_1\, T_1\, H^u$, we find that this condition also leads to two other new couplings: $T_1\,M_1\,H^d$ and $H^d\,M_1\,\nu^c_1$. In all, the structure of the superpotential in this model (continuing to suppress coefficients) is,
\begin{align}
W=&\,\,\,\,\,\,T_1\,T_1\,H^u\,\,+\,\,T_2\,T_3\,H^u\,\,+\,\,T_1\,M_1\,H^d\,\,+\,\,T_2\,M_3\,H^d\,\,+\,\,T_3\,M_2\,H^d\,\,+\,\,H^u\,M_1\,\nu _1^c\,\,\nonumber\\&\!\!+H^u\,M_2\,\nu _3^c\,\,+\,\,H^u\,M_3\,\nu _2^c\,\,+\,\,X_1\,Y_1\,Y_1^c\,\,+\,\,X_1\,Y_2\,Y_2^c\,\,+\,\,X_1\,N_1^c\,\nu _1^c+\,\,X_2\,N_1^c\,\nu _2^c\nonumber\\&\!\!+X_2\,N_2^c\,N_3^c.\label{fthw1}
\end{align}

This model has all the ingredients of the `Diamond Ring' proposed in \mbox{Ref.\ \cite{Heckman:2008ads}} and more. In particular, it can have an up-type quark mass matrix which is dominated by a single eigenvalue and a generic-enough set of down-type quark masses. Notice also that (\ref{fthw1}) includes a rich neutrino sector, with both Dirac-type masses of the form $H^uM_i\,\nu_j^c$ and Majorana-like masses between the fields $\nu_i^c$ and the Standard Model-singlet $N_1^c$. And if the singlet fields $X_i$ were to develop large vacuum expectation values and supersymmetry breaking F-terms---as argued in \cite{Heckman:2008ads}---the messengers \mbox{$Y_1\oplus Y_1^c$} and \mbox{$Y_2\oplus Y_2^c$} would communicate this to the Standard Model through gauge interactions. 

A full analysis of this model would be beyond the scope of this pedagogical discussion, but the structure of (\ref{fthw1}) is, at first glance, remarkably realistic. The interested reader should consult \mbox{Ref.\ \cite{Heckman:2008ads}} for more details. But unlike in \cite{Heckman:2008ads}, because every interaction in our model is generated locally within a {\it single, explicit patch} of a fibration, we have complete control over the local moduli relevant to the physics of (\ref{fthw1}). These are the moduli which determine the maps $a,c,$ and $d$ in the fibration (\ref{e8tosu5pinch}). If we take each of these maps to be linear functions of the two complex coordinates of the F-theory base $W$, then these correspond to $3\times2$ coefficients and $3$ constants, making a total of nine complex parameters which specify the model\footnote{This is an over-counting of the actual free-parameters: we can shift the constants of each map to be relative to their `center of mass,' removing one complex parameter, and we can re-scale each of the coordinates of the base, removing another two coefficients. Relating equivalent `configurations' of singularities will further reduce the number of parameters.}.

\begin{figure}[!h]\begin{center}\includegraphics[height=14.85cm, width=11.45cm,angle=90]{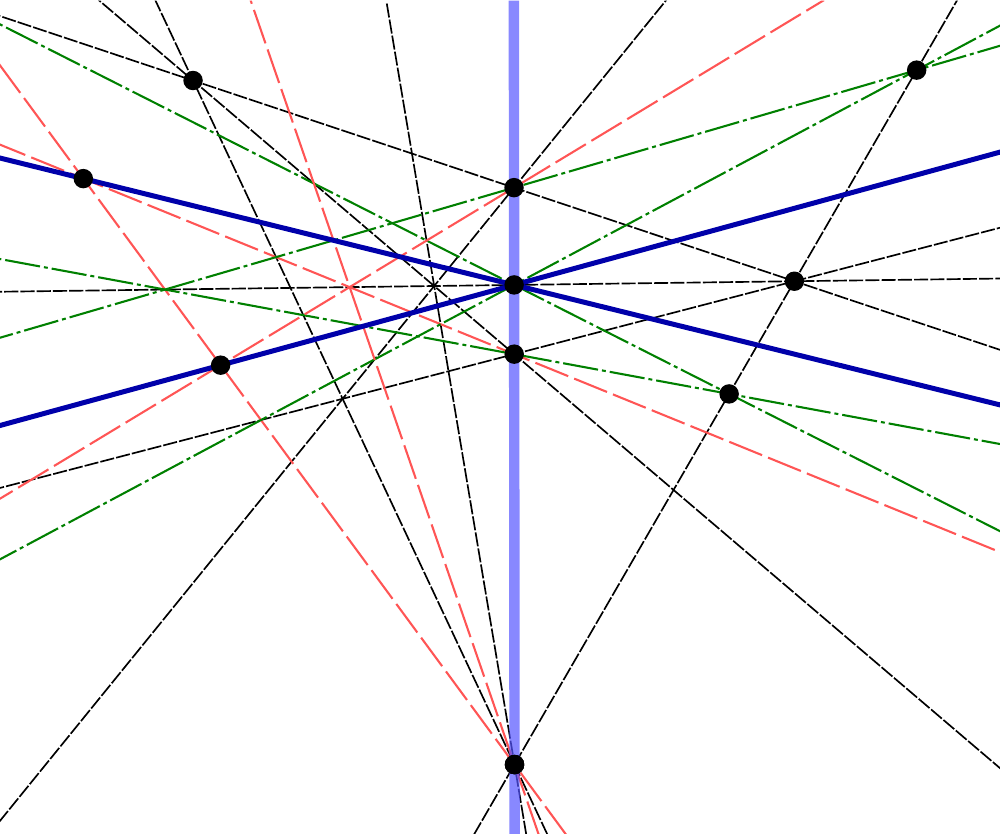}\caption{A randomly-chosen example of the fibration (\ref{e8tosu5pinch}). The thick, semi-transparent blue line indicates the two-rank enhanced singularity of type $SO_{12}$ along which is supported $T_1$ and $H^u$. Other curves supporting $\mathbf{10}$'s are drawn in thick, solid blue; those supporting the messenger fields $Y_i$ and $Y_i^c$ are in dash-dotted green while all other $\mathbf{5}$-dimensional representations are in widely-dashed red; $SU_5$-singlets are drawn in finely-dashed black. The gauge-invariant operators appearing in (\ref{fthw1}) are indicted by dots.\label{fthsu5model2slice}}\end{center}\vspace{-2cm}\end{figure}

Picking each of these nine parameters randomly, we obtain a picture of the fibration that is illustrated in \mbox{Figure \ref{fthsu5model2slice}}.

\newpage

\subsubsection{The Solitaire Diamond Ring of F-Theory}\label{solitaire}
\begin{figure}[h]\begin{center}\includegraphics[width=13cm,height=4.2cm,angle=0]{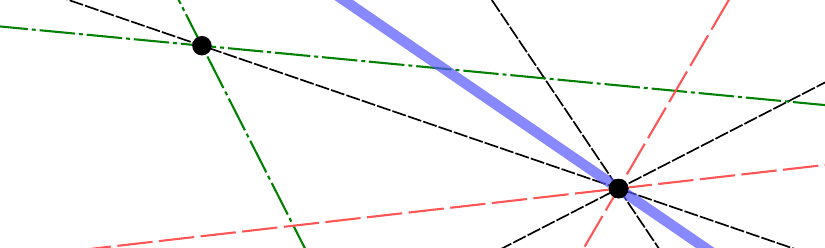}\caption{A realization of the fibration (\ref{e8tosu5pinch}), drawing those matter-curves with non-vanishing flux, as listed in \mbox{Table \ref{fthsu5model3}}. The curves are coloured as in \mbox{Figure \ref{fthsu5model2slice}}. \label{fthsu5model32slice}}\end{center}\vspace{-1.6cm}\end{figure}

One of the possibilities in F-theory that we have not yet exploited is that each matter-curve can support any number of generations, depending on the amount of $U_1$-flux turned on along it. Therefore, we can choose to support all three generations of matter along the doubly-enhanced $SO_{12}$-curve $-c-d=0$ of \mbox{Table \ref{fthsu5model2}}. By doing this, we reduce the number of matter-curves necessary to achieve a complete model. An example of such a realization is given in \mbox{Table \ref{fthsu5model3}}, and is illustrated in \mbox{Figure \ref{fthsu5model32slice}}.

Like above, this model achieves all the structure described by Ref.\ \cite{Heckman:2008ads} and more; but unlike our last example, this model does so with only nine matter-curves. As such, this realization is closer in spirit to the original `Diamond Ring' of Ref.\ \cite{Heckman:2008ads}, but here includes also an interesting neutrino sector. In all, the superpotential generated in F-theory for this fibration and choice of fluxes is of the form, \vspace{-0.15cm}
\begin{equation}
\hspace{-0.0cm}W=\,\,\lambda_1^{ij} T_i\,T_j\,H^u\,\,+\,\,\lambda_2^{ij}T_i\,M_j\,H^d\,\,+\,\,\lambda_3^{ij}H^u\,M_i\,\nu_j^c\,\,+\,\,\lambda_4X\,Y\,Y^c\,\,+\,\,\lambda_5^iX\,\nu_i^c\,N^c.\vspace{-0.15cm}
\end{equation}

As before, it would be beyond the scope of our current discussion to investigate the physics resulting from this model in any detail; we refer the reader to \mbox{Ref.\ \cite{Heckman:2008ads}}. Our purpose here is simply to show the plausibility of such a model by constructing it explicitly in F-theory as a local $\widehat{E_8}$-fibration. 

\newcommand{\FtheoryThT}[6]{$#1$&$\mathbf{#2}$&\ifnum#3<0$\!\!$#3\else#3\fi&\ifnum#4<0$\!\!$#4\else#4\fi&\ifnum#5<0$\!\!$#5\else#5\fi&$#6=0$\\}
\begin{table}[h]\begin{center}
\vspace{-0.2cm}\begin{tabular}{|rllll|r|}
\hline
$\,\qquad\!\!$&$\!\!\!SU_{5}\times\,\!\!\!\!$&$\!U_1^a\times\!\!\!\!\!$&$\!U_1^b\times\!\!\!\!\!$&$\!U_1^c$&Location\\\hline
\FtheoryThT{3\times\,\,T_i}{10}{0}{-1}{-1}{-c - d}
\FtheoryThT{3\times M_i}{\bar{5}}{0}{-1}{3}{-c + 3 d}
\FtheoryThT{H^u\!}{5}{0}{2}{2}{2 c + 2 d}
\FtheoryThT{H^d\!}{\bar{5}}{0}{2}{-2}{2 c - 2 d}
\FtheoryThT{Y}{5}{-2}{-2}{2}{-2 a - 2 c + 2 d}
\FtheoryThT{Y^c\!}{\bar{5}}{2}{-2}{-2}{2 a - 2 c - 2 d}
\FtheoryThT{X\,}{1}{0}{4}{0}{4 c}
\FtheoryThT{3\times\,\,\nu_i^c}{1}{0}{-1}{-5}{-c - 5 d}
\FtheoryThT{N^c}{1}{0}{-3}{5}{-3 c + 5 d}
 \hline
\end{tabular}\caption{The matter resulting from the fibration (\ref{e8tosu5pinch}), subject to a choice of fluxes chosen to minimize the number of matter-curves necessary to achieve realistic phenomenology.\label{fthsu5model3}}\end{center}\vspace{-0.6cm}\end{table}

\newpage~
\subsection{Phenomenological Models in M-Theory}\label{mtheg}
Unlike in F-theory, in M-theory we cannot simply ignore unwanted chiral matter if their corresponding singularities are present in the fibration (see \mbox{Section \ref{fmdiffs2}}). This means that additional exotic matter, such as that on the right-hand side of \mbox{Table \ref{fthsu5modelall}} from \mbox{Section \ref{fthsu5canon}}, will be more problematic in M-theory. Without dynamics to lift unwanted matter out of the low-energy effective theory, the only sure way to avoid exotics is to force their corresponding singularities out of the local geometry altogether through constraints on the maps defining the fibration. This virtual necessity makes models in M-theory much more difficult to maneuver, and very much more constrained---and hence predictive---than their F-theory cousins. 

But perhaps more problematic than exotic matter, any purely local model in M-theory with at least $SU_5$ gauge symmetry will suffer from the doublet-triplet splitting problem---the problem that unless the coloured partners of the Standard Model Higgs fields are `split' away from their partners and projected-out at a very high-scale, they will ordinarily lead to rapid proton decay. The F-theory solution proposed in \mbox{Refs.\ \cite{Beasley:2008dc,Beasley:2008kw}} does not have any known perturbative analogue in M-theory, and the M-theory mechanism described by \mbox{Ref.\ \cite{Witten:2001bf}} requires detailed knowledge of the compact geometry, which is inaccessible from the local perspective. Because doublet-triplet splitting is critical to the stability of the proton, we are led to more drastic (hence predictively constrained) options in M-theory. 

The only clear way to `solve' the doublet-triplet splitting problem locally in M-theory, then, is to work in a framework where the doublet and triplet components of the $\mathbf{5}$-dimensional Higgs fields are manifestly distinguished\footnote{An alternative approach could be to construct a model based on what is known as flipped $SU_5$, where the breaking of $SU_5$ by the vevs of $\mathbf{10}\oplus\bar{\mathbf{10}}$ (the exotics of \mbox{Table \ref{fthsu5modelall}}) automatically generates masses for the Higgs colour-triplets. See, e.g., \cite{Antoniadis:1987dx,Antoniadis:1989zy}; for a realization of flipped $SU_5$ in the context of Dbrane models, see \cite{Axenides:2003hs,Kokorelis:2008ce}. 
}. Specifically, if we begin with a model having only \mbox{$SU_3\times SU_2\times U_1^Y$} gauge symmetry, then because the Higgs fields and their coloured-partners are manifestly distinguished with respect to $SU_3\times SU_2$, it may be possible either to dynamically project-out the colour-triplet fields themselves, or to simply forbid the operators dangerous to proton decay without spoiling the ordinary Higgs-sector of the theory. The example we discuss below exploits both of these options.

\newcommand{\MTheoryZero}[9]{\vspace{-0.02cm}$#1$&$\mathbf{#2}$&$\mathbf{#3}$&\ifnum#4<0$\!\!$#4\else#4\fi&\ifnum#5<0$\!\!$#5\else#5\fi&\ifnum#6<0$\!\!$#6\else#6\fi&\ifnum#7<0$\!\!$#7\else#7\fi&\ifnum#8<0$\!\!$#8\else#8\fi&$#9=0$\\}

\subsubsection{The M-Theory Minimal Supersymmetric Standard Model}\label{mtheorymssm}
Our starting point for building a high-scale MSSM in M-theory is the fibration (\ref{e8tosu5resolution}) from \mbox{Section \ref{fthsu5canon}} which describes a general $\widehat{E_8}$-fibration giving $SU_5$ gauge symmetry:\vspace{-0.2cm}
\begin{equation}\widehat{E_8}(a+b+c+d,a-b+c+d,-c-d,-c-d,-c-d,-c-d,-c+3d,2c-2d).\label{e8tosu5resolutionprime}\vspace{-0.2cm}\end{equation}
The complete set of matter-generating singularities present in the geometry when $a,b,c,$ and $d$ are general maps was listed in \mbox{Table \ref{fthsu5modelall}}.

Our first challenge is to unfold the fibration (\ref{e8tosu5resolutionprime}) into one where only $SU_3$ and $SU_2$ singularities are present in the typical fibre. And we should do this in such a way so as to make hypercharge manifest. In units where the hypercharge of each quark doublet is $+1$, we are led to the following parameterization of the $\widehat{E_8}$-fibration,
\begin{align}\hspace{-1.05cm}\widehat{E_8}&\left(a+b+c+d+\tfrac{2}{3}Y,\right. a-b+c+d+\tfrac{2}{3}Y,-c-d-\tfrac{7}{3}Y,-c-d-\tfrac{7}{3}Y,\nonumber\\&\qquad\qquad\qquad\left. -c-d+\tfrac{8}{3}Y,-c-d+\tfrac{8}{3}Y,-c+3d-\tfrac{4}{3}Y,2c-2d-\tfrac{4}{3}Y\right).\label{mtheory_fibration1}
\end{align}
To check that this works, notice that all the generically-shrunk two-cycles in this fibration are generated by the cycles labelled by\vspace{-0.5cm}\begin{equation}\left\{(e_5-e_6)\,\,(-e_0+e_6+e_7+e_8)\right\}\otimes\left\{(e_3-e_4)\right\},\vspace{-0.3cm}\end{equation}
which are the simple roots of $SU_3$ and $SU_2$, respectively. 

Going through the now-familiar exercise of determining the matter content in such a fibration, we find those fields listed in \mbox{Tables \ref{MtheoryMSSM_project} and \ref{MtheoryMSSM_keep}}. The matter listed in \mbox{Table \ref{MtheoryMSSM_project}} are the (unwanted) `exotic' fields---which would be generally present in the fibration (\ref{mtheory_fibration1}) if $a,b,c,d$ and $Y$ were all generic maps. Notice that in addition to the descendants of the right-hand side of \mbox{Table \ref{fthsu5modelall}}, \mbox{Table \ref{MtheoryMSSM_project}} also includes the exotic (anti) quark doublet $Q_Y$ descending from the adjoint of $SU_5$,\vspace{-0.3cm}\begin{equation}\mathbf{24}=(\mathbf{8},\mathbf{1})_0\oplus(\mathbf{1},\mathbf{3})_0\oplus(\mathbf{3},\mathbf{2})_{\text{-}5}\oplus(\bar{\mathbf{3}},\mathbf{2})_{5}.\vspace{-0.3cm}\end{equation} This field lives at the solution to $5Y(t)=0$, and so the the easiest way to exclude it from the spectrum is to require that $Y(t)$ be a non-vanishing constant, independent of the base coordinate $t\in W$. Similarly, it is easy to see that a sufficient condition for excluding {\it all} the matter in \mbox{Table \ref{MtheoryMSSM_project}} is that the maps $c$ and $d$ also be non-vanishing constants, independent of the base. Making this restriction on the space of maps will result in a fibration where only singularities corresponding to the matter listed in \mbox{Table \ref{MtheoryMSSM_keep}} are present. 

\begin{table}[h]\begin{center}
\begin{tabular}{|llllllll|r|}
\hline
$\,\qquad\!\!$&$\!\!\!SU_3\times\,\!\!\!\!$&$\!\!\!SU_2\times\!\!\!\!\!$&$\!U_1^a\times\!\!\!\!\!$&$\!U_1^b\times\!\!\!\!\!$&$\!U_1^c\times\!\!\!\!\!$&$\!U_1^d\times\!\!\!\!\!$&$\!U_1^Y$&Location\\\hline
\MTheoryZero{Q_X}{3}{2}{0}{0}{0}{4}{1}{4 d + Y}
\MTheoryZero{Q_X^c}{\bar{3}}{2}{0}{0}{-3}{1}{-1}{-3 c + d - Y}
\MTheoryZero{Q_Y}{\bar{3}}{2}{0}{0}{0}{0}{5}{5 Y}
\MTheoryZero{D_X}{3}{1}{0}{0}{-3}{-3}{-2}{-3 c - 3 d - 2 Y}
\MTheoryZero{U_X^c}{\bar{3}}{1}{0}{0}{0}{4}{-4}{4 d - 4 Y}
\MTheoryZero{U_X}{3}{1}{0}{0}{-3}{1}{4}{-3 c + d + 4 Y}
\MTheoryZero{L_X}{1}{2}{0}{0}{-3}{-3}{3}{-3 c - 3 d + 3 Y}
\MTheoryZero{E_X^c}{1}{1}{0}{0}{0}{4}{6}{4 d + 6 Y}
\MTheoryZero{E_X}{1}{1}{0}{0}{-3}{1}{-6}{-3 c + d - 6 Y}
\MTheoryZero{\nu_X}{1}{1}{0}{0}{-3}{5}{0}{-3 c + 5 d}\hline
\end{tabular}\caption{Exotic matter singularities that would typically be present in the fibration (\ref{mtheory_fibration1}) for generic maps $a,b,c,d,$ and $Y$. Notice that all these fields are independent of $a$ and $b$.\label{MtheoryMSSM_project}}\vspace{-3cm}\end{center}\end{table}

\newpage~
\vspace{2.3cm}
\begin{table}[h]\begin{center}\vspace{-3cm}\caption{The collection of matter fields present in the fibration (\ref{mtheory_fibration1}), excluding those listed in \mbox{Table \ref{MtheoryMSSM_project}}. These correspond to the locations of conical singularities over the base when the maps $c,d,$ and $Y$ are taken to be (generic) non-vanishing constants.\vspace{0.2cm}\label{MtheoryMSSM_keep}}
\begin{tabular}{|llllllll|r|}\hline$\,\qquad\!\!$&$\!\!\!SU_3\times\,\!\!\!\!$&$\!\!\!SU_2\times\!\!\!\!\!$&$\!U_1^a\times\!\!\!\!\!$&$\!U_1^b\times\!\!\!\!\!$&$\!U_1^c\times\!\!\!\!\!$&$\!U_1^d\times\!\!\!\!\!$&$\!U_1^Y$&Location\\\hline\MTheoryZero{Q_1}{3}{2}{1}{1}{-1}{-1}{1}{a + b - c - d + Y}
\MTheoryZero{Q_2}{3}{2}{1}{-1}{-1}{-1}{1}{a - b - c - d + Y}
\MTheoryZero{Q_3}{3}{2}{-2}{0}{-1}{-1}{1}{-2 a - c - d + Y}
\MTheoryZero{u_1^c}{\bar{3}}{1}{1}{1}{-1}{-1}{-4}{a + b - c - d - 4 Y}
\MTheoryZero{u_2^c}{\bar{3}}{1}{1}{-1}{-1}{-1}{-4}{a - b - c - d - 4 Y}
\MTheoryZero{u_3^c}{\bar{3}}{1}{-2}{0}{-1}{-1}{-4}{-2 a - c - d - 4 Y}
\MTheoryZero{d_1^c}{\bar{3}}{1}{1}{1}{-1}{3}{2}{a + b - c + 3 d + 2 Y}
\MTheoryZero{d_2^c}{\bar{3}}{1}{1}{-1}{-1}{3}{2}{a - b - c + 3 d + 2 Y}
\MTheoryZero{d_3^c}{\bar{3}}{1}{-2}{0}{-1}{3}{2}{-2 a - c + 3 d + 2 Y}
\MTheoryZero{D_1}{3}{1}{-1}{-1}{-2}{2}{-2}{-a - b - 2 c + 2 d - 2 Y}
\MTheoryZero{D_2}{3}{1}{-1}{1}{-2}{2}{-2}{-a + b - 2 c + 2 d - 2 Y}
\MTheoryZero{D_3}{3}{1}{2}{0}{-2}{2}{-2}{2 a - 2 c + 2 d - 2 Y}
\MTheoryZero{D_1^c}{\bar{3}}{1}{-1}{-1}{-2}{-2}{2}{-a - b - 2 c - 2 d + 2 Y}
\MTheoryZero{D_2^c}{\bar{3}}{1}{-1}{1}{-2}{-2}{2}{-a + b - 2 c - 2 d + 2 Y}
\MTheoryZero{D_3^c}{\bar{3}}{1}{2}{0}{-2}{-2}{2}{2 a - 2 c - 2 d + 2 Y}
\MTheoryZero{L_1}{1}{2}{1}{1}{-1}{3}{-3}{a + b - c + 3 d - 3 Y}
\MTheoryZero{L_2}{1}{2}{1}{-1}{-1}{3}{-3}{a - b - c + 3 d - 3 Y}
\MTheoryZero{L_3}{1}{2}{-2}{0}{-1}{3}{-3}{-2 a - c + 3 d - 3 Y}
\MTheoryZero{H_1^u}{1}{2}{1}{1}{2}{2}{3}{a + b + 2 c + 2 d + 3 Y}
\MTheoryZero{H_2^u}{1}{2}{1}{-1}{2}{2}{3}{a - b + 2 c + 2 d + 3 Y}
\MTheoryZero{H_3^u}{1}{2}{-2}{0}{2}{2}{3}{-2 a + 2 c + 2 d + 3 Y}
\MTheoryZero{H_1^d}{1}{2}{1}{1}{2}{-2}{-3}{a + b + 2 c - 2 d - 3 Y}
\MTheoryZero{H_2^d}{1}{2}{1}{-1}{2}{-2}{-3}{a - b + 2 c - 2 d - 3 Y}
\MTheoryZero{H_3^d}{1}{2}{-2}{0}{2}{-2}{-3}{-2 a + 2 c - 2 d - 3 Y}
\MTheoryZero{e_1^c}{1}{1}{1}{1}{-1}{-1}{6}{a + b - c - d + 6 Y}
\MTheoryZero{e_2^c}{1}{1}{1}{-1}{-1}{-1}{6}{a - b - c - d + 6 Y}
\MTheoryZero{e_3^c}{1}{1}{-2}{0}{-1}{-1}{6}{-2 a - c - d + 6 Y}
\MTheoryZero{\nu_1^c}{1}{1}{1}{1}{-1}{-5}{0}{a + b - c - 5 d}
\MTheoryZero{\nu_2^c}{1}{1}{1}{-1}{-1}{-5}{0}{a - b - c - 5 d}
\MTheoryZero{\nu_3^c}{1}{1}{-2}{0}{-1}{-5}{0}{-2 a - c - 5 d}
\MTheoryZero{S_1}{1}{1}{-1}{-1}{4}{0}{0}{-a - b + 4 c}
\MTheoryZero{S_2}{1}{1}{-1}{1}{4}{0}{0}{-a + b + 4 c}
\MTheoryZero{S_3}{1}{1}{2}{0}{4}{0}{0}{2 a + 4 c}
\MTheoryZero{N_1}{1}{1}{-3}{1}{0}{0}{0}{-3 a + b}
\MTheoryZero{N_2}{1}{1}{0}{-2}{0}{0}{0}{-2 b}
\MTheoryZero{N_3}{1}{1}{3}{1}{0}{0}{0}{3 a + b}\hline
\end{tabular}\end{center}\vspace{0.0cm}\end{table}

\begin{figure}[b]\begin{center}\vspace{-13.5cm}\mbox{\hspace{-0.85cm}\includegraphics[scale=1.1,angle=89.5]{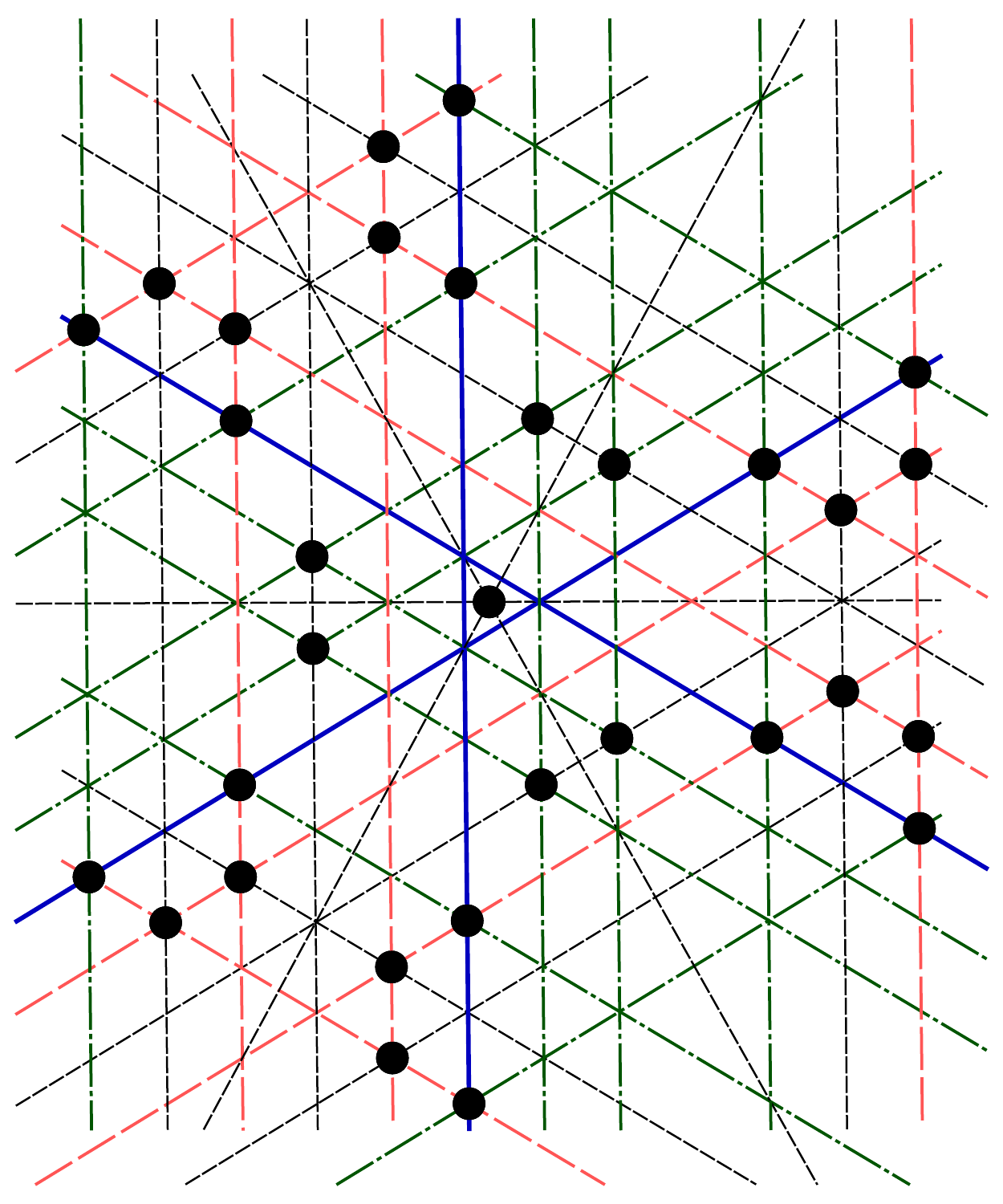}}\\\vspace{-0.3cm}\mbox{\hspace{4.cm}$H^d_2\hspace{0.8cm} \nu^c_2\hspace{0.8cm} u^c_2\hspace{0.2cm} N_2\hspace{1.4cm} D^c_2\hspace{0.8cm} D_2\hspace{0.8cm}Q_2\hspace{0.8cm} L_2\hspace{0.8cm} S_2 $}\vspace{-0.2cm}\caption{A particular F-theory realization of the fibration (\ref{mtheory_fibration1}), imposing the `parallelization-constraints' necessary to project-out the unwanted matter content of \mbox{Table \ref{MtheoryMSSM_project}}, leaving only that of \mbox{Table \ref{MtheoryMSSM_keep}}. Quark bi-fundamentals are coloured in thick, solid blue, $SU_3$-triplets are in dash-dotted green, electroweak doublets are in widely-dashed red, and non-Abelian singlets are drawn in finely-dashed black. Each dot represents a non-vanishing cubic term in the superpotential\label{mtheoryeg1}. Each matter-curve is labelled as in \mbox{Table \ref{MtheoryMSSM_keep}}}\vspace{-1.55cm}\end{center}\vspace{-0.5cm}\end{figure}

By restricting $c,d,$ and $Y$ to be constant maps, we greatly reduce the moduli space of the resulting fibration. Indeed, if we rescale and translate the coordinate $t$ so that $a(t)=t$, the entire fibration can specified by only ten real numbers: one from the coefficient of $t$ in $b(t)$, and nine from the four constant three-vectors in the maps $b,c,d,$ and $Y$---considering $O_3$-related configurations to be equivalent. Because there are only two non-constant maps, any F-theory realization of this fibration will possess a great deal of `accidental' discrete flavour symmetry among the matter-curves. One such realization is shown in \mbox{Figure \ref{mtheoryeg1}}, where this accidental symmetry is manifestly visible. In M-theory, this discrete symmetry of the F-theory plane would not generically be present---to see this, imagine using any random slice through \mbox{Figure \ref{mtheoryeg1}} to define a fibration for M-theory.\vspace{0.2cm}

\hspace{4.35cm}$d^c_1\hspace{0.9cm} e^c_1 \hspace{0.9cm} H^u_1 \hspace{0.8cm} S_1 \hspace{0.8cm} L_1 \hspace{0.8cm} Q_1\hspace{0.8cm} D_1\hspace{0.8cm} D^c_1\hspace{0.65cm} $\vspace{-0.25cm}

\vspace{0.85cm}\raisebox{-5cm}{$\hspace{-1.95cm}\begin{array}{c}H^d_3\vspace{0.45cm}\\\nu^c_3\\N_1\vspace{0.05cm}\\u^c_3\vspace{1.45cm}\\D^c_3\vspace{0.45cm}\\D_3\vspace{0.5cm}\\Q_3\vspace{0.5cm}\\L_3\vspace{0.5cm}\\S_3\vspace{0.5cm}\\H^u_3\vspace{0.5cm}\\e^c_3\vspace{0.5cm}\\d^c_3\vspace{-2cm}\end{array}\vspace{-2cm}$}\vspace{-11.15cm}

\hspace{15.05cm}\mbox{$\begin{array}{c}\vspace{0.4cm}\\u^c_1\vspace{0.85cm}\\N_3\vspace{0.2cm}\\\nu^c_1\vspace{1.6cm}\\H^d_1\vspace{3.2cm}\\\!\!\!d^c_2\vspace{1.55cm}\\\!\!\!e^c_2\vspace{1.6cm}\\\!\!\!H^u_2\vspace{-3cm}\end{array}$}\vspace{-12cm}

\newpage

The matter content of \mbox{Table \ref{MtheoryMSSM_keep}} should be reminiscent of that which descends from three $\mathbf{27}$'s of an $E_6$ grand-unified model; however, there are some important differences. Under ordinary group-theoretic branching\footnote{For a review of $E_6$-like grand-unified models, see, e.g., \cite{Hewett:1988xc}.}, the cubic interaction among three $\mathbf{27}$'s would descend to a large number of operators dangerous to proton stability, such as the di-quark interactions \vspace{-0.2cm}\begin{equation}Q\,Q\,D\quad\mathrm{and}\quad D^c\,u^c\,d^c,\vspace{-0.2cm}\end{equation} and the lepto-quark interactions \vspace{-0.2cm}\begin{equation}Q\,D^c\,L,\quad D\,d^c\,\nu^c,\quad\mathrm{and}\quad D\,u^c\,e^c\vspace{-0.2cm}\end{equation}---where we have denoted the Higgs' colour-triplet partners as $D$ and $D^c$. But in M-theory (as in F-theory), we have the freedom to choose the relative conjugations of fields living at distinct conical singularities independently\footnote{Notice that this is remarkably more restrictive than the corresponding freedom in F-theory. In F-theory, one can obtain along a given matter-curve any integer $n$ of left-handed fields transforming under a particular representation, where $n$ can be positive, negative, or vanishing. In M-theory, we have the freedom to choose only $n=\pm1$.} \cite{Bourjaily:2009aa}. In particular, this allows us to exclude {\it all} these dangerous couplings of the colour-triplet fields simply by choosing their conjugations opposite to that which descends from $\mathbf{27}$'s---choosing instead to take those triplets descending from $\bar{\mathbf{27}}$'s. Such a choice was made when preparing \mbox{Table \ref{MtheoryMSSM_keep}}.

We have also made use of this freedom to conjugate the $SO_{10}$-singlet components of the $\mathbf{27}$'s, denoted $S_i$ in \mbox{Table \ref{MtheoryMSSM_keep}}. This allows us to retain interactions of the form $S\,D\,D^c$ while simultaneously excluding those of the form $S\,H^u\,H^d$. Notice that this means that the operator $S^{\dag} H^u H^d$ {\it is} gauge-invariant, and can potentially lead to a Giudice-Masiero-like \cite{Giudice:1988yz}, dynamically-generated $\mu$-term from the K\"{a}hler potential, while the operator $S\,D\,D^c$ in the superpotential can generate masses for the Higgs colour-triplets. 

In all, the cubic-level superpotential generated in this theory (suppressing coefficients) is as follows:
\begin{align}
\hspace{-0.75cm}W_3=&\,\,\,\,\,Q_1\,u_2^c\,H_3^u\,\,+\,\,Q_1\,u_3^c\,H_2^u\,\,+\,\,Q_2\,u_3^c\,H_1^u\,\,+\,\,Q_2\,u_1^c\,H_3^u\,\,+\,\,Q_3\,u_1^c\,H_2^u\,\,+\,\,Q_3\,u_2^c\,H_1^u\nonumber
\\&\!\!+Q_1\,d_2^c\,H_3^d\,\,+\,\,Q_1\,d_3^c\,H_2^d\,\,+\,\,Q_2\,d_3^c\,H_1^d\,\,+\,\,Q_2\,d_1^c\,H_3^d\,\,+\,\,Q_3\,d_1^c\,H_2^d
\,\,+\,\,Q_3\,d_2^c\,H_1^d\nonumber
\\&\!\!+D_1\,D_2^c\,S_3\,\,+\,\,D_1\,D_3^c\,S_2\,\,+\,\,D_2\,D_3^c\,S_1\,\,+\,\,D_2\,D_1^c\,S_3\,\,+\,\,D_3\,D_1^c\,S_2\,\,+\,\,D_3\,D_2^c\,S_1
\nonumber\\&\!\!+L_1\,e_2^c\,H_3^d\,\,+\,\,L_1\,e_3^c\,H_2^d\,\,+\,\,L_2\,e_3^c\,H_1^d\,\,+\,\,L_2\,e_1^c\,H_3^d\,\,+\,\,L_3\,e_1^c\,H_2^d\,\,+\,\,L_3\,e_2^c\,H_1^d\nonumber
\\&\!\!+L_1\,\nu _2^c\,H_3^u\,\,+\,\,L_1\,\nu _3^c\,H_2^u\,\,+\,\,L_2\,\nu _3^c\,H_1^u\,\,+\,\,L_2\,\nu _1^c\,H_3^u
\,\,+\,\,L_3\nu_1^c\,H_2^u\,\,+\,\,L_3\nu_2^c\,H_1^u
\nonumber\\&\!\!+N_1\,N_2\,N_3.\label{MthW}
\end{align}


\noindent Although somewhat beyond the scope of our present analysis, it is worth mentioning that even at the level of supersymmetric effective field theory, it is possible to observe dynamics that can generate large vacuum expectation values for some or more of the Standard Model-singlet fields $S_i$, $N_i$, or $\nu^c_i$. If we include the FI-parameters for each of the (locally-)anomalous\footnote{Actually, $U_1^d$ is locally non-anomalous. This is easy to see, because, relative to the traditional branching of three $\mathbf{27}$'s, we have only conjugated $U_1^d$-charged fields in pairs. It is not clear what this implies about the viability of $U_1^d$ as a gauge symmetry, but it means that we should not include an FI-parameter for $U_1^d$ in its D-term equation.} $U_1$-symmetries corresponding to $a,b,c,$ and $d$, it is possible for their D-term equations to require large vevs for some or all of these fields\footnote{Indeed, using the cubic superpotential in (\ref{MthW}) and generic FI-parameters, the F- and D-term equations would generate vevs for $N_1$, $S_1$, $S_3$, and $\nu_1^c$. Nonetheless, we prefer to keep our discussion here general, because, for example, which fields get vevs can easily change as a result of conjugating some of the Standard Model-singlets. It is interesting to note that only slight modifications must be made to our choice of conjugations in order to achieve gauge coupling unification, while preserving our conclusions about quark masses and proton stability.}, similar to as in \cite{Blumenhagen:2006xt,Choi:2006bh}). This would have the effect of projecting out some of the colour-triplets at a high scale, possibly leaving some in the low-energy spectrum to accommodate gauge-coupling unification\footnote{The absence of lepto-quark and di-quark interactions in (\ref{MthW}) is sufficient for proton longevity, regardless of the scale at which the colour-triplets are projected out of the spectrum; this will be explained below.}. Of course, because we have started with a high-scale $SU_3\times SU_2\times U_1^Y$ model, gauge coupling unification would be motivated by low-scale phenomenology rather than strictly imposed by high-scale unification. 

In addition to projecting out the Higgs-triplets, large singlet vacuum expectation values could justify initially including more of the exotic matter listed in \mbox{Table \ref{MtheoryMSSM_project}}, relaxing the constraints we had to impose on the fibration (\ref{mtheory_fibration1}). For example, it may be possible to start with a model in M-theory with `$(4-1)$'-generations, where a vector-like pair of generations is dynamically projected-out at a very-high sale. This idea turns out to be quite tenable, and will be explored in greater detail in \cite{Bourjaily:2009ab}. 

But if Standard Model-singlet fields acquire large-enough vevs, quartic operators in the superpotential involving these singlets could become effective cubic operators, possibly relevant to low-scale physics. In our example model, with the singularities listed in \mbox{Table \ref{MtheoryMSSM_keep}}, there are $129$ three-cycles which generate quartic operators---too many to justify listing them all here; and all but fifteen of these operators involve Standard Model-singlets that could possibly acquire large vevs. Therefore, there can be a great deal of variability in the structure of the {\it effective} superpotential, depending on how many (and which) singlet fields get vevs.  Among these quartic operators, for example, are 
\begin{equation}\begin{split}W_4\supset&\,\,\,\,\,\,\, \eta_1^u\,N_2\,Q_1\,u_1^c\,H_3^u\,\,+\,\, \eta_2^u\,N_1\,Q_2\,u_2^c\,H_1^u\,\,+\,\, \eta_3^u\,N_3\,Q_3\,u_3^c\,H_2^u\\
&+\eta_1^d\,N_2\,Q_1\,d_1^c\,H_3^d\,\,+\,\,\eta_2^d\,N_1\,Q_2\,d_2^c\,H_1^d\,\,+\,\,\eta_3^d\,N_3\,Q_3\,d_3^c\,H_2^d,\label{mthegquart}\end{split}\vspace{-0.3cm}\end{equation}
where \begin{equation}\eta_i^{u,d}\sim\frac{1}{\Lambda_{i}^{u,d}}e^{-\mathrm{Vol}(\Sigma_{i}^{u,d})}\,;\end{equation}
here, each $\Lambda_i$ is set by the length-scale of the corresponding tree-cycle volume in Planck units. Such a suppression can be surmounted if one of the singlets $N_i$ were to acquire a large-enough vacuum expectation value through D-term equations involving FI-parameters. Because the only term in the superpotential (\ref{MthW}) that involves the singlets $N_i$ is the operator $N_1\,N_2\,N_3$, F-term equations will allow at most one of the fields $N_i$ to acquire a vev without spoiling supersymmetry. Because such a vev can in principle be quite large (set by the scale of the FI-parameters\footnote{These parameters are loop-suppressed relative to the Planck scale (see, e.g., \cite{Choi:2006bh}), but should be larger than the scales associated with local cycles that suppress quartic operators, which can in principle be as low as $M_{GUT}$.}), one quark of each type could develop an $\mathcal{O}(1)$-Yukawa coupling---and so we would expect no more than one parametrically heavy generation. Notice that many of the phenomenologically desirable interactions that were absent form the cubic superpotential can nonetheless be generated dynamically in the theory.

But we should worry that perhaps some of the effective operators generated by singlet-vevs could spoil, for example, the baryon- and lepton-number conservation of the cubic superpotential (\ref{MthW}), leading to rapid proton decay. For example, among the quartic couplings in this model are the following, 
\begin{equation}W_4\supset \frac{\lambda_1}{\Lambda_1}S_3\,D_2^c\,u_2^c\,d_3^c\,\,+\,\,\frac{\lambda_2}{\Lambda_2}S_3\,D_1\,u_3^c\,e_1^c\,,\label{mthquartpdecay}\end{equation}which violate baryon- and lepton-number, respectively. Because the vacuum expectation value of $S_3$ is that which generates an effective mass for the colour-triplets $D_1,\, D_2^c$ (see (\ref{MthW})), there appears to be competition between the need to project-out the dangerous triplet fields by giving $S_3$ a large vev, and the need to avoid dangerous effective operators such as those in (\ref{mthquartpdecay}). Integrating-out the field $S_3$, the quartic interactions in (\ref{mthquartpdecay}) would facilitate proton decay through the following effective tree-level process:
\vspace{-0.5cm}\begin{figure}[!h]\begin{center}\includegraphics[scale=1]{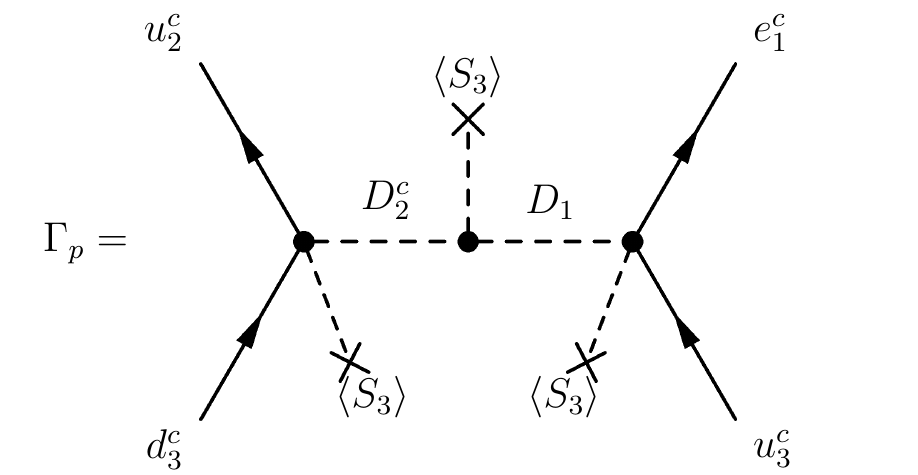}\end{center}\end{figure}\vspace{-0.75cm}

It is not hard to estimate the strength of this process. Let us denote the vev of $S_3$ by $V\equiv\langle S_3\rangle$, and the effective mass-term for $D_1,\,D_2^c$ by $\lambda V$. If we assume that all the three-cycle volumes are roughly the same, so $\lambda_1\sim\lambda_2\sim\lambda$---which implies that $\Lambda_1\sim\Lambda_2\sim\Lambda$---then we find that the rate scales roughly as
\begin{equation}\Gamma_p\sim\frac{\lambda V}{\Lambda}\frac{1}{\lambda^2 V^2}\frac{\lambda V}{\Lambda}\sim\frac{1}{\Lambda^2}.\label{pdecay}\end{equation}

Notice that this rate is insensitive to the vev of $S_3$. In particular, this means that whether the colour-triplets are projected out at $M_{GUT}$ or at $1\,\mathrm{TeV}$ makes almost no difference on the longevity of the proton. Rather, the nucleon decay rate (\ref{pdecay}) is set primarily by the scale $\Lambda$ corresponding to the size of local three-cycles in the geometry\footnote{Because the coefficients $\lambda_i$ are exponential in three-cycle volumes, taking their ratios to be $\sim1$ as we have done in equation (\ref{pdecay}) is rather na\"{i}ve. When doing the calculation correctly, we expect that these ratios of exponential coefficients $\lambda_i$ will be very significant to the analysis.}. Because these local three-cycles must be smaller than the three-cycle of the base of the fibration, they are bounded below by $M_{GUT}< \Lambda$,\footnote{Recall that the size of the three-cycle base of the fibration is set by the relation $\mathrm{Vol}(W)\sim g_{YM}^{-2}$ in Planck units.}. Because the decay rate is suppressed by a scale larger than $M_{GUT}$, we expect our model to be marginally safe from present experimental bounds\footnote{For background on proton decay in grand unified models, see, e.g., \cite{Dimopoulos:1981dw,Dimopoulos:1981zb,Ibanez:1981yh}; for a more recent discussion, see, e.g., \cite{Babu:1998ep,Murayama:2001ur}.}.

Although it would go beyond the scope of this note to study the viability of this model in any more detail, what we have seen so far is very encouraging. It may very well be that this model fails for a number of reasons upon detailed inspection, but it certainly establishes the plausibility of truly realistic, local, phenomenological models in M-theory. \\~\\

%

\section{Discussion and Future Directions}
The example models we have presented in this paper all feature a geometric analogue of traditional grand unification in the spirit of Refs.\ \cite{Bourjaily:2007vw,Bourjaily:2007vx,Bourjaily:2007kv}. What we mean is that there are directions in the {\it local} moduli-spaces of our fibrations where the gauge symmetry of the model can be smoothly enhanced, causing several disparate matter singularities to coalesce into those generating unified matter multiplets. 

For example, consider the moduli which locally deform the geometry of the $G_2$-manifold resulting from the $\widehat{E_8}$-fibration (\ref{mtheory_fibration1}) described in \mbox{Section \ref{mtheg}}. These moduli are the constants and coefficients of the maps $a,b,c,d,$ and $Y$. Locally, there would be no obstruction to smoothly sending the (constant) three-vector $Y$ to zero, for example. Upon doing this, the $SU_3$ and $SU_2$ co-dimension four singularities would coalesce into one of type $SU_5$, and the conical singularities giving rise to the $SU_3\times SU_2$ charged matter listed in \mbox{Tables \ref{MtheoryMSSM_project} and \ref{MtheoryMSSM_keep}} will coalesce into those listed in \mbox{Table \ref{fthsu5model2}}. This should not be surprising: the constant map $Y$ can be understood as controlling the `parallel' separation\footnote{But care must be taken with this picture of `parallel' stacks of branes. There is an {\it ultra}-local sense in which we can envision the $SU_3$ and $SU_2$ singularities in M-theory to be the fibre-wise duals of parallely-separated, flat D6-branes in type IIa, but this picture breaks down over the entire patch. Ultra-locally, we may envision, for example, a $(\mathbf{3},\mathbf{1})$ and a $(\mathbf{1},\mathbf{2})$ living along a single D6-brane which intersects two parallel stacks of D6-branes---a stack of three and two branes, respectively; the two stacks of branes can be smoothly brought into coincidence, and when this happens, the two originally disparate fields will coalesce into a single $\mathbf{5}$ of $SU_5$. But this picture breaks down more broadly simply because M-theory on an $\widehat{E_8}$-fibred manifold {\it doesn't have any perturbative type IIa dual.} This merely reflects the fact that $\widehat{E_8}$ does not possess any circle isometry upon which we can compactify M-theory. Another way to see this is by the fact that the stacks of branes in the complete fibration cannot be {\it flat}: there are {\it three} places where the $SU_2$-stack and the $SU_3$ stack intersect to support the three left-handed quark doublets in \mbox{Table \ref{MtheoryMSSM_keep}}.} between stacks of D6-branes in a dual type IIa picture. 

Of course, we actively engineered the fibration (\ref{mtheory_fibration1}) to have this structure by {\it unfolding} it out of one giving $SU_5$ gauge theory. And if we continue to reverse-engineer our work in \mbox{Section \ref{exempliModels}}, it is easy to see that sending $Y\to0$ and $d\to0$ simultaneously will result in a model with $SO_{10}$ gauge theory; and sending $c\to0$ as well will restore our starting place, a local $E_6$ grand unified model with three generations. But these are not the only corners of moduli space that exhibit some degree of grand unification. Indeed, almost all the `breaking' patterns from $E_6$ to the Standard Model are realizable. For example, in the limit where $Y\to d$, the $SU_3$-singularity giving rise to colour will become enhanced to $SU_4$, resulting in a Pati-Salam-like model with $SU_4\times SU_2$ gauge symmetry \cite{Pati:1973uk}.

But there are some important differences between this local, {\it geometric} analogue of unification and traditional grand-unification. First, there is little sense in which the matter representations of an unfolded model will become unified at high energies. Although our example of a geometry giving $SU_3\times SU_2\times U_1^Y$ gauge theory in \mbox{Section \ref{mtheg}} is smoothly related to one giving $SU_5$ gauge symmetry, the unfolded geometry of the fibration (\ref{mtheory_fibration1}) for $Y\neq0$ generates specifically $SU_3\times SU_2\times U_1^Y$ symmetry {\it at the high scale}. And in a compact model, the myriad apparently continuously-adjustable moduli such as $Y$ would be fixed by quantization conditions of the compactification. This is not dissimilar from theories of intersecting D-branes in type II string theory: locally, a stack of D-branes in flat space can be pulled apart, smoothly breaking $SU_5$ to $SU_3\times SU_2$, for example; but in a compact manifold, D-branes must wrap minimal-volume (supersymmetric) cycles, and these cycles are usually without any flat directions that would allow such parallel separations. 

Nonetheless, the specific values of the moduli parameters in the maps $a,b,c,d,$ and $Y$ will determine the extent to which the theory will {\it appear to be unified} from the low-energy point of view. Specifically, smoothly adjusting the moduli can continuously alter the degree to which the theory obeys particular features of traditional grand unification, such as Yukawa unification and mass relations. As one adjusts these local moduli, a wide variety of different structures emerge, from approximately-enhanced gauge symmetries to discrete flavour symmetries. And importantly, these structures do not need to `break' or disappear through a cascade of Higgs-mechanisms. Although there is a very concrete sense in which our realization of F-theory's `Diamond Ring' in \mbox{Section \ref{diamondring}} is locally, geometrically related to an $SO_{10}$ grand-unified model, this need not be realized at any energy scale, or even be a structure that is realizable in the global geometry. This allows us to make use of many of the important ideas of grand unification without committing ourselves to a cascading tower of effective field theories, requiring new symmetry breaking mechanisms at each stage. \\

In this paper, we have presented a proof of concept that {\it concrete, purely-local, semi-realistic phenomenological models with three generations of matter exist in F-theory and M-theory}. And we have seen how the variety of mutually interacting matter fields necessary to describe the Standard Model can only be realized at the very limit of complexity achievable in ALE-fibrations. What we loose in global constraints by considering such local models, we gain in concrete and highly-predictive structure---making it reasonable to expect that this framework of purely local models in string theory can be tested and falsified by experiment.

\acknowledgments
This work has been encouraged and supported by many people. The author is especially grateful for insightful discussions with Malcolm Perry, Eric Kuflik, Edward Witten, Paul Langacker, Herman Verlinde, Bobby Acharya, Gordy Kane, Stuart Raby, Cumrun Vafa, Jonathan Heckman, Nima Arkani-Hamed, and Gary Shiu. Jihye Seo made many valuable comments on early drafts of this paper, helping to clarify the places that needed clarification. The author is grateful for the hospitality of the Michigan Center for Theoretical Physics, the 2008 Simons Workshop on Mathematics and Physics, and IAS at the Hebrew University of Jerusalem. This work was supported in part by a Graduate Research Fellowship from the National Science Foundation.

\newpage~

\appendix
\section{ALE-Manfiolds and Their Moduli Spaces}\label{ALEApp}
Asymptotically Locally Euclidean (ALE) manifolds play an important role in string theory. We begin this appendix with a brief review of Felix Klein's classification of co-dimension four Calabi-Yau orbifolds from the 1880's \cite{Klein:1884}---largely for historic curiosity and as a general foundation for the more modern discussion to follow. Building upon Klein's work, we describe how these ADE-singularities can be resolved into smooth ALE-manifolds via complex structure deformations such as those used to geometrically engineer models in F-theory. Of course, complex structure deformations require that a complex structure be chosen for the ALE-space; but because ALE-manifolds are {\it hyper}-K\"{a}hler, they posses an entire $SU_2$-triplet of complex structures, and any particular choice will obscure some of the underlying geometry. 

Perhaps the best understanding of the ALE moduli-spaces in their full hyper-K\"{a}hler generality---as is necessary for building ALE-fibred $G_2$-manifolds in M-theory---is due to Kronheimer \cite{Kronheimer:1989zs}. We review his construction in \mbox{Appendix \ref{kornheimersconstruction}}, and conclude with a discussion of how Kronheimer's framework is related to the presentation of ALE moduli described by Katz and Morrison in \cite{Katz:1992ab}, which is the source of notation used throughout this paper.

\subsection{ADE Classification of Co-Dimension Four Calabi-Yau Orbifolds}
Co-dimension four orbifold-singularities can be written in the form $\mathbb{R}^4/\Gamma$ where $\Gamma$ is a discrete subgroup of $SO_4\equiv SU_2^L\times SU_2^R$---the rotations of the three-sphere at infinity. When $\Gamma$ is restricted to be a subset of only $SU_2^L$, for example, $\mathbb{R}^4/\Gamma$ is resolvable into a space having only $SU_2^L$ holonomy---which is therefore Calabi-Yau and hence also hyper-K\"{a}hler. And so, the co-dimension four Calabi-Yau orbifolds are classified simply by the discrete subgroups $\Gamma\subset SU_2$. 

Considering $SU_2$ as the double cover of $SO_3$, these are just the discrete symmetries among points on a two-sphere. These come in 
three varieties:
\begin{enumerate}
\item[{\bf A.}] $\mathbb{Z}_n\equiv\Gamma_{SU_n}(\equiv\Gamma_{A_{n-1}})$, the symmetries of a regular $n$-gon lying along the equator;
\item[{\bf D.}] $D_{n-4}\equiv\Gamma_{SO_{2n}}(\equiv\Gamma_{D_n})$, the dihedral symmetry group of a regular $n$-gon prism;
\item[{\bf E.}] and three exceptional subgroups, the symmetries of the five Platonic solids:
\begin{itemize}
\item $\Gamma_{E_6}$, the symmetries of the points of a tetrahedron (which is self-dual);
\item $\Gamma_{E_7}$, the symmetries of the points of a cube or its dual, the octahedron;
\item $\Gamma_{E_8}$, the symmetries of the points of an icosahedron or its dual, the dodecahedron.
\end{itemize}
\end{enumerate}
\newpage
\subsubsection{Felix Klein and the Icosahedron}
This classification of Calabi-Yau co-dimension four orbifolds dates as far back as Felix Klein's {\it Lectures on the Icosahedron}, published in 1884 \cite{Klein:1884}, where the icosahedron in the title is obviously a reference to $\Gamma_{E_8}$,\footnote{Of course, Klein himself did not use this notation: his lectures predate the classification of Lie algebras by several years.}.

After classifying the singularities as we have just done, Klein constructed sets of $\Gamma_{ADE}$-invariant variables in $\mathbb{C}^2/\Gamma_{ADE}$, any three of which must be related to one another by a single equation; this allows for $\mathbb{C}^2/\Gamma_{ADE}$ to be described isomorphically as a hypersurface in $\mathbb{C}^3$---the set of solutions to the equation relating the three $\Gamma_{ADE}$-invariant variables. 

Take for example the orbifold $\mathbb{C}^2/\Gamma_{SU_n}$, where the action of $\Gamma_{SU_n}=\mathbb{Z}_n$ on $(a,b)\in\mathbb{C}^2$ is generated by $\Gamma_{SU_n}\ni \eta:(a,b)\mapsto(\eta a, \eta^{-1}b)$ with $\eta^n=1$. It is not difficult to find $\Gamma_{SU_n}$-invariant variables: for example, let $x\equiv a^n$, $y\equiv b^n$, and $z\equiv ab$; these are all $\Gamma_{SU_n}$-invariant and are related by the equation $xy=z^n$, the standard hypersurface realization of the $SU_n$-orbifold. 

Notice that the asymptotic structure of the hypersurface \mbox{$f(x,y,z)=xy-z^n=0$} is unchanged if we add any $(n-1)^{\mathrm{th}}$-degree polynomial of $z$ to $f(x,y,z)$; and if this polynomial has non-vanishing discriminant, the modified hypersurface will be non-singular. In modern terminology, adding such terms to $f$ amounts to a complex structure deformation of the hypersurface, whose generally-smooth limit is the $\widehat{SU_n}$ ALE-space.

Through similar, yet considerably more involved analyses, Klein also constructed hypersurfaces isomorphic to each of the other ADE-orbifolds. The complete list is given in Table \ref{orbifolds}; when presented as such, the ADE-orbifolds are commonly referred to as `Kleinian singularities.'
 
\begin{table}[!h]\begin{center}
\vspace{-0.3cm}\begin{tabular}{lcr}
\hline
Orbifold&{~}&Hypersurface\\
\hline $SU_n$ ($\equiv A_{n-1}$) &{~\qquad\qquad~}& $xy=z^n$\\
$SO_{2n}$ ($\equiv D_{n}$) && $x^2+y^2z=z^{n-1}$\\
$E_6$ && $x^2=y^3+z^4$\\
$E_7$ && $x^2+y^3=16yz^3$\\
$E_8$ && $x^2+y^3=z^5$\\\hline
\end{tabular}\caption{\label{orbifolds}Kleinian singularities as hypersurfaces in $\mathbb{C}^3$.}\vspace{-0.75cm}\end{center}\end{table}

It is a straight-foward (if rather difficult) exercise in algebraic geometry to construct the complete set of complex structure deformations consistent with the asymptotic symmetries of each ADE-orbifold in \mbox{Table \ref{orbifolds}}. The general deformations of each hypersurface up to and including $\widehat{E_7}$ were worked out by Bramble\footnote{Albeit with a few typos, corrected by \cite{Katz:1992ab}. In Bramble's defense, however, it should be noted that even the presentation of the deformations of $\widehat{E_7}$ given in \cite{Katz:1992ab}---which spans several pages of an appendix---is not without its own typographical errors (although none that are substantive).} as early as 1918 \cite{Bramble:1918}, but those of $\widehat{E_8}$ had to wait until 1992 when Katz and Morrison implicitly described the general form of its deformations in \cite{Katz:1992ab}.

Among other results in \cite{Katz:1992ab}, the authors describe a general procedure by which any particular presentation of a resolved $\widehat{E_n}$ can be converted into a canonical form, expressed in terms of functions on an especially convenient basis of moduli parameters. Just for the sake of concreteness, their canonical form of a resolution of $\widehat{E_8}$ as a function of its eight complex structure moduli $(f_1,\ldots,f_8)$ is given by the curve \begin{equation}x^2+y^3-z^5+\epsilon_{2}\,y\,z^3+ \epsilon_8\, y\, z^2+ \epsilon_{12}\, z^3+ \epsilon_{14}\, y\, z+ \epsilon_{18}\,z^2+ \epsilon_{20}\, y+\epsilon_{24}\,z+\epsilon_{30}=0,\label{e8res}\end{equation}where $\epsilon_{n}$ is an $n^{\mathrm{th}}$-degree symmetric function of the moduli $(f_1,\ldots,f_8)$. Although explicit forms of the coefficients $\epsilon_{n}$ have still not appeared in the literature, it is unlikely that anyone will ever care to publish them: after working through the machinery in \cite{Katz:1992ab}, one finds that at least seventeen pages would be required to write them out completely in standard form\footnote{For the reader who is interested in working with the full form of equation (\ref{e8res}), the author of this note would be happy to supply a Mathematica file containing it.}.

For us, however, the important point of Katz and Morrison's analysis, is not that they (implicitly) give the complete form of equation (\ref{e8res}); but rather, that by canonically organizing the deformation moduli according to the roots of $E_8$, the geometry of $\widehat{E_8}(f_1,\ldots,f_8)$ can be analyzed without making any reference whatsoever to equation (\ref{e8res}). The map between the moduli parameters $(f_1,\ldots,f_8)$ and the geometry of $\widehat{E_8}$ was given in Table \ref{En_roots_and_areas}, and is reproduced below.\\
\vspace{3.5cm}

\begin{table}[h]\vspace{-0.3cm}
\begin{tabular}{rcc@{}c}
\cline{2-4}
&$\begin{array}{c}\text{Positive Roots of~}E_n,\\\text{Labels~of~Two-Cycles~in~}\widehat{E_n}\end{array}$&&$\begin{array}{c}\text{`Area' of Corresponding}\\\text{Two-Cycle in~}\widehat{E_n}(f_1,\ldots,f_n)\end{array}$\\
\cline{2-4}& $e_i-e_{j>i}$&$\implies$&$f_i-f_{j>i}$\\
&$-e_0+e_i+e_j+e_k$&$\implies$& $f_i+f_j+f_k$\\
\footnotesize{$n\geq 6$}&$-2e_0+\sum_{j=1}^6e_{i_j}$ &$\implies$& $\sum_{j=1}^6f_{i_j}$\\
\footnotesize{$n=8$}&$-3e_0+e_{i}+\sum_{j=1}^8e_{j}$&$\implies$& $f_i+\sum_{j=1}^8f_j$\\
\cline{2-4}
\end{tabular}\caption{\label{En_roots_and_areas2}The roots of $E_n$---written as vectors in $\mathbb{R}^{n,1}$ having squared-norm of 2---which are in one-to-one correspondence with two-cycles in $\widehat{E_n}$ whose areas are fixed by the moduli $(f_1,\ldots,f_n)$. This is adapted from Table 4 of \mbox{Ref.\ \cite{Katz:1992ab}}. }\vspace{-0.2cm}\end{table}

\newpage
\subsection{Kronheimer's Construction of ALE-Manifolds and Their Moduli}\label{kornheimersconstruction}\vspace{-0.3cm}
Fortunately, however, there is a dramatically easier way of understanding the moduli space of deformations of ADE-singularities  that keeps manifest both the hyper-K\"{a}hler structure of the resolutions and the origin of the moduli in terms of the root-lattices of ADE-groups. This is due to Kronheimer \mbox{\cite{Kronheimer:1989zs}}, and represents his response to the challenge set by Hitchin in the classic paper \cite{Hitchin:1994zr} discussing the $SU_n$-singularities and their resolutions in complete hyper-K\"{a}hler detail. What Kronheimer showed in \mbox{Ref.\ \cite{Kronheimer:1989zs}} is that the complete ALE-resolution of any particular rank-$n$ ADE-singularity can be described as the vacuum manifold of the $\mathcal{N}=2$ quiver gauge theory whose quiver is nothing other than the corresponding ADE-group's (extended) Dynkin diagram, with each node labelled by its {\it Dynkin index} $d_i$---contributing the gauge-group factor $U_{d_i}=SU_{d_i}\times U_1$ to the quiver. 

Because each node contributes a $U_1$-factor\footnote{The extended node is an overall, irrelevant $U_1$ that plays no role in the analysis.}, the vacuum of this quiver is fixed only after specifying the $n$ generalized FI-parameters\footnote{In $\mathcal{N}=2$, the D-term equation for each vector multiplet is related by an $SU_2^R$ R-symmetry to the F-term equations of its chiral component field, making the familiar $\mathcal{N}=1$ FI-parameter generalize to an entire $SU_2^R$-triplet of \mbox{FI-parameters}.}, $\left\{\zeta_i\right\}$. These are nothing but the full hyper-K\"{a}hler deformation moduli of the ADE-singularity's ALE-resolution.
\begin{figure}[h]\begin{center}\mbox{\hspace{-0.5cm}\includegraphics[scale=1.]{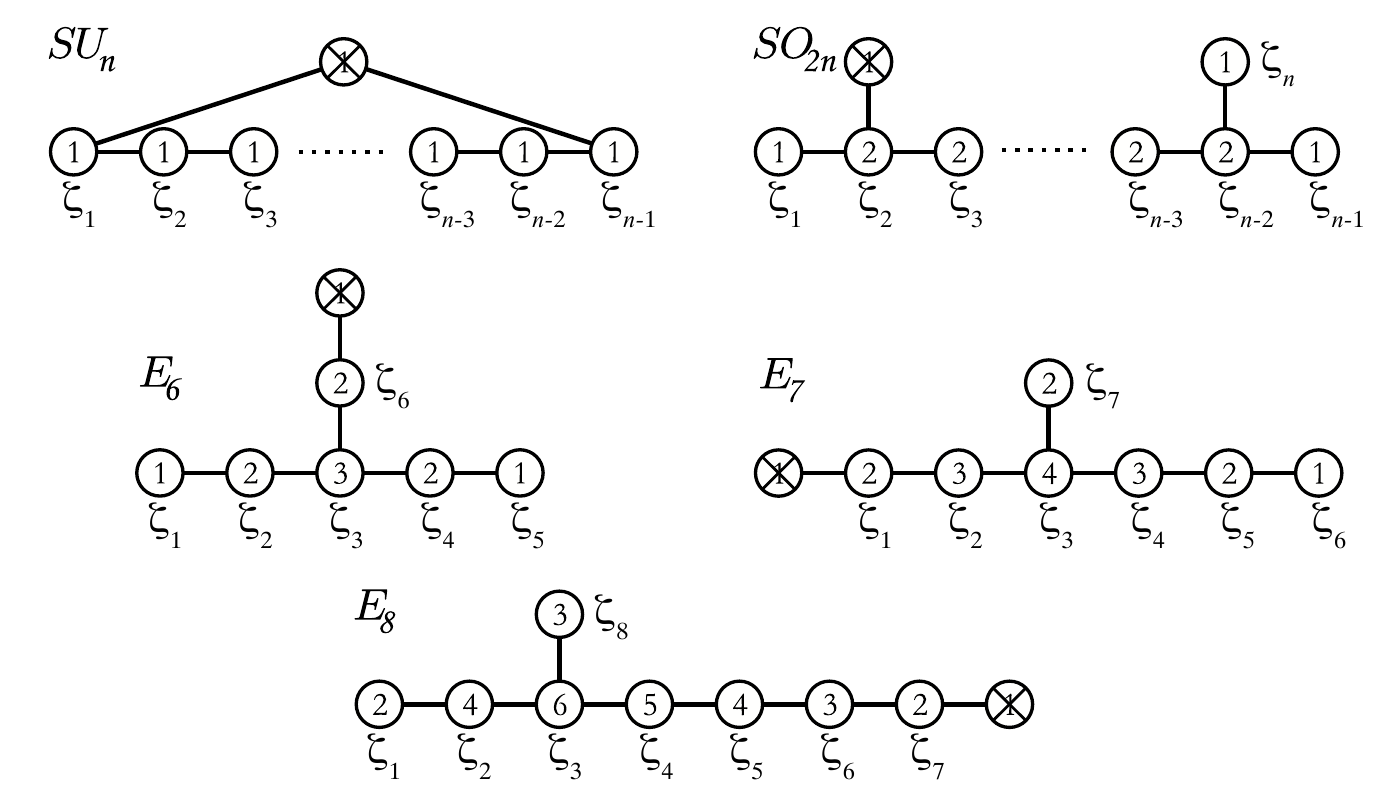}}\caption{Kronheimer showed that each ALE-space can be given as the vacuum manifold of the $\mathcal{N}=2$ quiver gauge theory whose quiver is the Dynkin-index-labelled (extended) Dynkin diagram of the corresponding ADE-group. The moduli space of a rank-$n$ ALE-space is then given by the $n$ generalized FI-parameters which specify the vacuum of the quiver gauge theory. See, e.g., \cite{Lindstrom:1983rt} for background on $\mathcal{N}=2$ quiver gauge theories. \label{kron}}\end{center}\end{figure}

\hspace{-0.2cm}To see how this works, consider the quiver gauge theory given on\hspace{0.25cm}\raisebox{-0.8cm}{\parbox{1.2cm}{\includegraphics[scale=1.25]{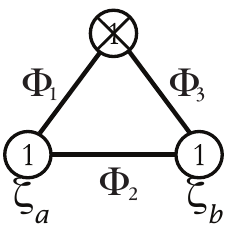}}}
\vspace{-2.cm}

\noindent the right, with the generalized FI-parameters $\vec{\zeta}_a$ and $\vec{\zeta}_b$. The vacuum\\ 
manifold of this quiver is
found by solving the F/D-term equations: 
\begin{align}
\vec{\zeta}_a&=\Phi_1^{\dag}\vec{\sigma}\Phi_1-\Phi_2^{\dag}\vec{\sigma}\Phi_2\equiv\vec{e}_1-\vec{e}_2=(1,-1,0);\nonumber\\
\vec{\zeta}_b&=\Phi_2^{\dag}\vec{\sigma}\Phi_2-\Phi_3^{\dag}\vec{\sigma}\Phi_3\equiv\vec{e}_2-\vec{e}_3=(0,1,-1).
\end{align}

Letting $(\varphi_i^+,\varphi_i^-)$ denote the complex scalar components of the hypermultiplet $\Phi_i$, we can construct the $(U_1^a\times U_1^b)$-invariant variables\begin{equation}x\equiv\prod_{i=1}^3\varphi_i^+,\qquad y\equiv\prod_{i=1}^3\varphi_i^-,\quad\mathrm{and}\quad z\equiv\varphi_1^+\varphi_1^-.\label{xyz}\end{equation} And fixing a complex structure, so that we can write \begin{equation}e_i\mapsto\left(\begin{array}{c}\mathfrak{Re}(\varphi_i^+\varphi_i^-)\\\mathfrak{Im}(\varphi_i^+\varphi_i^-)\\\left|\varphi_i^+\right|^2-\left|\varphi_i^-\right|^2\end{array}\right)\qquad\mathrm{and}\qquad \vec{\zeta}_{a,b}\equiv\left(\begin{array}{c}\mathfrak{Re}(f_{a,b})\\\mathfrak{Im}(f_{a,b})\\\rho_{a,b}\end{array}\right),\label{cstruct}\end{equation} we see that the variables of (\ref{xyz}) are related by \begin{equation}xy=z(z-f_a)(z-f_a-f_b).\label{su3klein}\end{equation} This equation describes a resolved $\widehat{SU_3}$ ALE-space. Notice that $\rho_a$ and $\rho_b$ in (\ref{cstruct}) play absolutely no role in (\ref{su3klein}), reflecting the loss of information associated with choosing a complex structure. 

Slightly less obvious, but reviewed in \cite{Hitchin:1994zr}, it turns out that there are three generally non-vanishing two-cycles in this space, only two of which are independent. They have the generalized `areas' (as measured by each of the three K\"{a}hler forms), \begin{equation}\int_{\Sigma_a}\vec{\omega}=\vec{\zeta}_a,\qquad\int_{\Sigma_b}\vec{\omega}=\vec{\zeta}_b,\qquad\int_{\Sigma_{a+b}}\vec{\omega}=\vec{\zeta}_a+\vec{\zeta}_b.\end{equation}

It is not difficult to generalize this simple example to arbitrary resolutions of $SU_n$-singularities (see, e.g., \cite{Acharya:2001gy,Berglund:2002hw,Acharya:2004qe}). But explicit demonstrations linking Kronheimer's construction to the complex structure deformations of the canonical Kleinian hypersurfaces $\widehat{SO_{2n}}$ and $\widehat{E_n}$ are considerably more challenging. The interested reader should consult Refs.\ \cite{Lindstrom:1999pz,Park:1999hu,Albertsson:2001jq}.
\newpage

From Kornheimer's construction we have a natural basis for the moduli---that corresponding to each of the generalized FI-parameters. This basis is very intuitive: referring to \mbox{Figure \ref{kron}}, if for $\widehat{E_8}$ one were to set all the FI-parameters to zero except $\zeta_2$, for example, then the vacuum manifold will be found to have $SU_7$ and $SU_2$-singularities separated by a two-cycle whose `area' is equal to $\zeta_2$. That is, turning on a single FI-parameter has the effect of blowing up the corresponding node in the geometry's Dynkin diagram-worth of two-cycles. And so each FI-parameter corresponds to a different rank-seven subgroup $H\times U_1\subset E_8$.

Nonetheless, when resolving $\widehat{E_n}$-spaces more generally, it is often more convenient to work in a basis that more closely reflects the underlying root-lattice structure. This is the symmetrical basis used by most researchers in geometric engineering since Katz and Morrison, including the present author. To convert between the two bases, one can use the following:

\begin{equation}f_i=\left(\mathcal{M}^n_{ij}\right)_{(\zeta\mapsto f)}\zeta_j,\qquad\mathrm{or}\qquad\zeta_j=\left(\mathcal{M}^n_{ij}\right)^{-1}_{(f\mapsto \zeta)}f_i,\end{equation}
where
\begin{equation}\hspace{-0.38cm}
\left(\mathcal{M}^n_{ij}\right)_{(\zeta\mapsto f)}=\left(\begin{array}{ccccccc}-\tfrac{2}{3}&-\tfrac{1}{3}&0&0&\cdots&0&\tfrac{1}{3}\\\tfrac{1}{3}&-\tfrac{1}{3}&0&0&\cdots&0&\tfrac{1}{3}\\\tfrac{1}{3}&\tfrac{2}{3}&1&0&\ddots&0&\tfrac{1}{3}\\\tfrac{1}{3}&\tfrac{2}{3}&1&1&\ddots&0&\tfrac{1}{3}\\\vdots&\vdots&\vdots&\ddots&\ddots&\vdots&\vdots\\\tfrac{1}{3}&\tfrac{2}{3}&1&1&\cdots&0&\tfrac{1}{3}\\\tfrac{1}{3}&\tfrac{2}{3}&1&1&\cdots&1&\tfrac{1}{3}\end{array}\right),\quad\mathrm{and}\quad\left(\mathcal{M}^n_{ij}\right)^{-1}_{(f\mapsto\zeta)}=\left(\begin{array}{cccccc}-1&1&0&0&\cdots&0\\0&-1&1&0&\ddots&0\\0&0&-1&1&\ddots&0\\\vdots&\ddots&\ddots&\ddots&\ddots&\vdots\\0&0&0&\cdots&-1&1\\1&1&1&0&\cdots&0\end{array}\right).
\end{equation}

\newpage
\section{Yukawa Couplings in F-Theory and M-Theory}
\subsection[Gauge-Invariant Operators and Triple-Intersections in F-Theory]{Correspondence Between Gauge-Invariant Cubic Operators\\ and Triple-Intersections in F-Theory}\label{TIeqGIO}
Consider a manifold that is described as a rank-$n$ ALE-space $\widehat{\mathfrak{g}}(f_1,\ldots,f_n)$ fibred over any base such that the typical fibre has an $\widehat{\mathfrak{h}}$-type singularity (where $\mathfrak{h}$ can be trivial). If the rank of $\mathfrak{h}$ is $(n-k)$, then the typical fibre will have $k$-independent two-cycles. By analogy to our work in \mbox{Section \ref{sectexempli}}, the complete set of matter curves in the fibration follows from the branching\begin{equation}\mathrm{adj}(\mathfrak{g})=\mathrm{adj}\left(\mathfrak{h}\times U_1^1\times\cdots\times U_1^k\right)\bigoplus_{\substack{\vec{q}=(q_1,\ldots,q_k)\\\mathrm{|~}\vec{q}\in Q^+}}\left\{\mathbf{R}_{(q_1,\ldots,q_k)}\oplus\bar{\mathbf{R}}_{(-q_1,\ldots,-q_k)}\right\},\label{adjdecomp}\end{equation}
where $Q^+$ is the set of charges $\vec{q}$ such that the first nonzero charge is positive. As discussed in \mbox{Section \ref{sectexempli}}, there is a `matter-curve' for each of the charges $\vec{q}$---where two proportional $\vec{q}\,$'s correspond to the same curve. We claim that whenever two distinct curves intersect, there exists a third that also intersects the pair. In view of (\ref{adjdecomp}), it is sufficient for us to show that whenever $\mathrm{adj}(\mathfrak{g})\supset\mathbf{R}_{\vec{q}_a}\oplus\mathbf{R}_{\vec{q}_b}$ for $\vec{q}_a\not\propto\vec{q}_b$, then also $\mathrm{adj}(\mathfrak{g})\supset\mathbf{R}_{\vec{q}_a\pm\vec{q}_b}$.

Suppose that the curves corresponding to $\vec{q}_a$ and $\vec{q}_b$ intersect. Because they intersect, they must be linearly independent, and so we may choose $\vec{q}_a$ and $\vec{q}_b$ to correspond to the curves $a=0$ and $b=0$, where $a\not\propto b$. Equivalently, we may use $\vec{q}_a$ and $\vec{q}_b$ to be distinct basis-elements of the space of $U_1$-charges. Now, in each fibre along the base, there are sets of two-cycles arranged in the weight-lattices of the representations $\mathbf{R}_{\vec{q}_a}$ and $\mathbf{R}_{\vec{q}_b}$ having areas $a$ and $b$, respectively. For the sake of simplicity, let us suppose that $\mathbf{R}_{\vec{q}_a}$ and $\mathbf{R}_{\vec{q}_b}$ are both irreducible representations of $\mathfrak{h}$,\footnote{If they are not irreducible, then the same argument will apply, replacing, for example, ``the highest-weight of $\mathbf{R}$'' with ``one of the highest-weights of $\mathbf{R}$,'' and similarly throughout.}.

Let $\alpha$ and $\beta$ denote the highest-weight two cycles of $\mathbf{R}_{\vec{q}_a}$ and $\mathbf{R}_{\vec{q}_b}$, respectively; and let us use $\left\{\eta\right\}$ to denote the simple roots of $\mathfrak{h}$. Adjoining the root $(\pm\alpha)\oplus\left\{\eta\right\}$ will generate an $(n-k+1)$-rank algebra, and similarly for the root $\pm\beta$,\footnote{The signs in $\pm\alpha,\pm\beta$, are to force consistent `positivity' of the roots $\alpha$, $\beta$, and those of $\left\{\eta\right\}$.}. Because the roots $\alpha$ and $\beta$ correspond to independent two-cycles in homology, they can simultaneously be adjoined to $\left\{\eta\right\}$ to form an $(n-k+2)$-rank algebra, which we will denote $\mathfrak{f}$,\footnote{The singularity at the point of intersection may not be (as small as) $\widehat{\mathfrak{f}}$---we have {\it not} required that $\alpha$ and $\beta$ are the {\it only} linearly-independent two-cycles that collapse at the intersection. This allows us to accommodate singularities that are multiply-enhanced in rank.} a subalgebra of the lattice of shrunk cycles at the intersection. We may without loss of generality assume that $\mathfrak{f}$ is a simple algebra, because $\alpha$ and $\beta$ must connect to the same subset of simple roots in $\left\{\eta\right\}$, or they would be separated in the fibres, and hence would not intersect.

The purpose of introducing $\mathfrak{f}$ is that it allows us to focus our attention on the simpler problem of the decomposition $\mathrm{adj}(\mathfrak{f})$ into $\mathfrak{h}\times U_1^a\times U_1^b$. Now, we claim that there is a set of weights in $\mathrm{adj}(\mathfrak{f})$ with $U_1^a\times U_1^b$-charges $\vec{q}_a\pm\vec{q}_b$. Because the two-cycles corresponding to such weights would have area $a\pm b$, the place over the base where this vanishes represents another matter-curve, distinct from $a=0$ and $b=0$, which supports matter in a representation with $U_1^a\times U_1^b$-charges $\vec{q}_a\pm\vec{q}_b$. It is not hard to construct such a weight: take the sum of all the weights $\pm\alpha\pm\beta\pm\bigoplus_i\eta_i$; this is a root in the weight-lattice of $\mathfrak{f}$, which has a corresponding two-cycle of area $a+b$, and shrinks to zero size when $a=b=0$. This weight is part of some representation $\mathbf{R}_{\vec{q}_a\pm\vec{q}_b}$, which has a corresponding curve distinct from, and intersecting with both curves $a=0$ and $b=0$.

\vspace{-0.6cm}  \begin{flushright} $~^{\text{`}}\!\!\acute{o}\pi\epsilon\rho\,\,~^{\text{'}}\!\!\acute{\epsilon}\delta\epsilon\iota\,\,\delta\epsilon
\!\!\text{\t{~}}\!\!\iota\xi\alpha\iota$\end{flushright}

\subsection{Collinearity of Conical Singularities in M-Theory}\label{collininM}
Suppose that the fields $A,B$, and $C$ arise from (isolated) conical singularities located at $t_a,t_b,$ and $t_c$ along the three-dimensional base $W$; what we mean is that there are sets of two-cycles in the fibration that have areas $f_a(t), f_b(t)$, and $f_c(t)$ (which depend linearly on the base), and that these maps vanish at $t_a,t_b,$ and $t_c$, respectively---that is, $f_{a}(t_a)=f_b(t_b)=f_c(t_c)=0$. We claim that if the operator $A\, B\, C$ is gauge-invariant, then $t_a,t_b,$ and $t_c$ are collinear in $W$.

Let \vspace{-0.4cm}\begin{equation}\begin{split}f_a(t)&=\alpha t-a_0;\\f_b(t)&=\beta t-b_0;\\f_c(t)&=\gamma t-c_0.\end{split}\end{equation}
Now, complete gauge-invariance implies $U_1$-invariance, and so $f_a(t)+f_b(t)+f_c(t)=0$. This means that $a_0+b_0+c_0=0$ and $\gamma=(-\alpha-\beta)$---and both are non-vanishing. From this it is straight forward to check that \begin{equation}t_c=\frac{a_0+b_0}{\alpha+\beta}=\frac{\beta}{\alpha+\beta}(t_b-t_a)+t_a;\end{equation} which is to say that $t_c$ lies along the line connecting $t_a$ and $t_b$ along the base. \vspace{-0.4cm} \begin{flushright} $~^{\text{`}}\!\!\acute{o}\pi\epsilon\rho\,\,~^{\text{'}}\!\!\acute{\epsilon}\delta\epsilon\iota\,\,\delta\epsilon
\!\!\text{\t{~}}\!\!\iota\xi\alpha\iota$\end{flushright}\vspace{-1cm}

\newpage


\end{document}